\documentclass[aps,prb,twocolumn,groupedaddress]{revtex4-1}
\usepackage{graphicx} 
\usepackage{dcolumn}
\usepackage{bm}
\usepackage{amsmath}
\usepackage{epsfig}
\usepackage{float}
\usepackage{epstopdf}
\usepackage{natbib}

\begin{document}
	\title{Lightwave Terahertz Quantum Manipulation of  Non-equilibrium Superconductor Phases  and their Collective Modes}%

	\author
	{M.~Mootz$^{1}$, J. Wang$^{2}$, and 
		I.~E.~Perakis$^{1}$ }
		\affiliation{$^1$ Department of Physics, University of Alabama at Birmingham, Birmingham, AL 35294-1170, USA \\
$^2$
		Department of Physics and Astronomy and Ames Laboratory-U.S. DOE, Iowa State University, Ames, Iowa 50011, USA.  }
	\date{\today}
	
	

\begin{abstract}
We present  a gauge-invariant density matrix 
description of  non-equilibrium superconductor (SC) states with spatial and temporal correlations driven  by  intense terahertz (THz) lightwaves.
We derive  superconductor Bloch--Maxwell equations of motion that extend  Anderson pseudo-spin models to  include the Cooper pair center-of-mass motion and  electromagnetic propagation effects. 
 We thus describe   
 quantum 
control of dynamical phases, collective modes, quasi-particle  coherence, 
and high nonlinearities  
during cycles of carrier wave oscillations, which 
relate to our recent experiments. 
Coherent photogeneration of a nonlinear  supercurrent  
with dc component via  condensate  acceleration 
by an effective lightwave field  dynamically  breaks the equilibrium inversion symmetry.    Experimental signatures  include 
  high harmonic light emission at equilibrium-symmetry-forbidden frequencies,  Rabi--Higgs collective modes and quasi-particle  coherence,  
  and 
   non-equilibrium  moving condensate states
   tuned by   few-cycle THz fields. 
     We use such lightwaves as an oscillating accelerating force that drives
     strong nonlinearities and anisotropic 
     quasi-particle populations 
   to control  and amplify different classes of collective  modes,  e.g., damped oscillations, persistent oscillations, and overdamped dynamics  via Rabi flopping.
  Recent phase-coherent nonlinear spectroscopy   experiments 
can  be modeled 
   by  solving the full nonlinear quantum dynamics  including self-consistent light--matter coupling.
\end{abstract}
	
	\maketitle

\section{Introduction}

Recent works have shown that  ultrafast phase-coherent THz nonlinear spectroscopy~\cite{yang2019lightwave,vaswani2019discovery,Luo2017,chu2020phase,Vaswani2020} is  a powerful tool for sensing and controlling non-equilibrium phases~\cite{mitrano,fausti,morris,porer,Li2013,lingos-17,
Shahbazyan2000,Shahbazyan2000b,josab,kondo,Yang2018b,yang2018} and collective modes~\cite{Pekker2015,shimano,leggett,krull,kemper}
 of  quantum materials. For example, 
 the non-equilibrium dynamics of quasi-particles (QPs) in SCs  has been characterized and controlled by THz pulses~\cite{Matsunaga:2013,Yang2019b}. THz  quantum quench of the SC order parameter 
by a single-cycle pulse   yields access to  a  long-lived (10's of ns)  gapless quantum fluid phase of QPs hidden by superconductivity~\cite{yang2018}. 
By tuning multi-cycle THz pulses, the above  QP  state  changes into 
 non-equilibrium gapless SC, i.~e., a 
moving condensate with gapless excitation spectrum, nearly unchanged macroscopic coherence, and infinite conductivity~\cite{yang2019lightwave,vaswani2019discovery}. 
In all  above  cases, the dynamics over 100's of ps is controlled 
by  lightwave few-cycle  fields that last for only few ps. 
Unlike photoexcitation  at optical frequencies, 
 THz lightwave electric fields   acts  as
oscillating  forces~\cite{langer,valley,huber-nat,Huber-subcycle} that accelerate the condensate and, in this way, 
control its excitation spectrum and order parameter as discussed here.

At the same time,  intense efforts have focused  on how to use ultrafast THz spectroscopy to detect the collective modes~\cite{Anderson,Nambu,Bardasis1961,Pekker2015,Littlewood1981,littlewoodb,varma,sherman}
that characterize  quantum phases and symmetry breaking in superconductors~\cite{shimano,matsunaga2014}.
 In  BCS superconductors, the electronic collective modes cannot be probed straightforwardly  with linear spectroscopy, as they require finite condensate momentum in order to couple to electromagnetic fields~\cite{Podolsky2011,Pekker2015}. If charge-density order coexists  with SC, the amplitude Higgs mode  becomes observable with Raman spectroscopy~\cite{Klein1980,Littlewood1981,littlewoodb}. Alternatively,  with dc supercurrent injection,  the Higgs mode can be detected with linear spectroscopy~\cite{Shimano2019,Moor2017}. Identifying amplitude  modes in the nonlinear response is possible via third harmonic generation in ultrafast THz spectroscopy, but this  is challenging because charge-density fluctuations dominate over the Higgs mode within BCS theory in a clean system~\cite{matsunaga2014,Aoki2017,udina}. Nevertheless, recent studies  argued that Higgs modes can still be observed if electron--phonon coupling or impurities are considered~\cite{Cea2018,Murotani}. Detection of purely electronic amplitude and phase collective modes and dynamical  phases in clean  superconductors remains an open challenge.

THz ultrafast spectroscopy experiments  have been mainly interpreted  so far in terms of  Anderson pseudo-spin precessions
 based on Liouville/Bloch equations and nonlinear response functions~\cite{Axt2007,Podolsky2011,Pekker2015,Aoki2017,Forster2017,udina}. 
However, THz  lightwave acceleration 
~\cite{langer,valley,huber-nat,Huber-subcycle} of the condensate 
during cycles of carrier wave oscillations 
and electromagnetic propagation effects 
have been mostly  neglected.
Recent experimental observations of high harmonic generation (HHG) at equilibrium-symmetry forbidden frequencies 
together with long-lived gapless quantum states~\cite{yang2019lightwave,vaswani2019discovery,yang2018} confirmed  the 
importance of  Cooper pair center-of-mass  momentum. However, such  
quantum transport effects 
during THz sub-cycle timescales~\cite{yang2019lightwave,vaswani2019discovery,langer,valley,huber-nat,Huber-subcycle}
 require an extension of  Anderson pseudo-spin precession models~\cite{Axt2007,Aoki2017,Forster2017,Wu2017,Wu2019}.

In this paper, we 
discuss  a  model for analyzing THz phase-coherent nonlinear spectroscopy experiments in quantum materials. This model is based on THz dynamical symmetry breaking during cycles of lightwave oscillations via lightwave condensate 
acceleration and electromagnetic propagation effects, as well as  high
Anderson pseudo-spin  nonlinearities. 
 For this, we 
 derive a gauge-invariant, non-adiabatic  density matrix theory 
for treating both temporal and spatial fluctuations 
in  combination with Maxwell's equations. Our theory extends previous studies of quantum transport~\cite{Stephen1965,Wu2017,haug} and HHG~\cite{Aoki2017,Cea2018} by including the nonlinear dynamics due to self-consistent light--matter electromagnetic coupling.
We use this theory to interpret recent experiments~\cite{yang2019lightwave,vaswani2019discovery,yang2018}
in terms of  THz dynamical  symmetry-breaking via nonlinear supercurrent coherent photogeneration.   We first present the full nonlinear quantum kinetic 
  theory, which treats spatial and temporal fluctuations, finite 
  Cooper pair center-of-mass condensate momentum $\mathbf{p}_\mathrm{S}(t)$, and  
 SC phase dynamics while observing gauge invariance. 
 We then apply a spatial gradient expansion of the full 
 equations that 
 allows the  separation  of the condensate center-of-mass and  Cooper pair relative motions
 analogous to the theory of ultrafast nonlinear quantum transport in semiconductors~\cite{haug}. We illustrate how a 
 dc nonlinear photocurrent component can be controlled by the cycles of oscillation 
 and the fluence of the pump pulse, as well as  by the thickness of a SC film.
 We also compare  the manifestations of charge-density fluctuations and collective modes in the highly nonlinear regime for a  one-band BCS model.  
  As a new application of  gauge invariant 
 non-perturbative treatment of  coupled  pseudo-spin precession, lightwave condensate acceleration, 
and  electrodynamics, 
 we demonstrate selective driving and control, 
by tuning few-cycle THz transient fields, 
 of different classes of  collective modes of the SC order parameter,  including damped oscillations, persistent (undamped) oscillations, overdamped dynamics,
amplified Higgs modes, etc. For example,  we demonstrate lightwave coherent control 
of all three dynamical phases predicted  theoretically by ``sudden quench"  of the SC order parameter~\cite{Yuzbashyan:2006}. We also show that 
Rabi--Higgs collective  modes can be driven by Rabi flopping, 
which modifies the nonlinear light emission spectrum. 
The strength of such Rabi--Higgs  oscillations is enhanced by the  interference between forward-moving  and reflected lightwaves 
inside a nonlinear SC thin film. 
Explicit  calculations of multi-dimensional THz coherent nonlinear spectra  will be presented elsewhere.
Here we make the point that the nonlinear interplay of Anderson pseudospin precession, Cooper pair quantum transport due to condensate 
lightwave acceleration, and electromagnetic field propagation effects 
must be treated self-consistently  in the time domain in order to interpret phase-coherent THz nonlinear spectroscopy experiments with  well-characterized, 
 phase-coherent, intense  THz pulses ~\cite{yang2019lightwave,vaswani2019discovery,yang2018,langer,valley,huber-nat,Huber-subcycle}.

\begin{figure}[!tbp]
	\centering
		\includegraphics[scale=0.23]{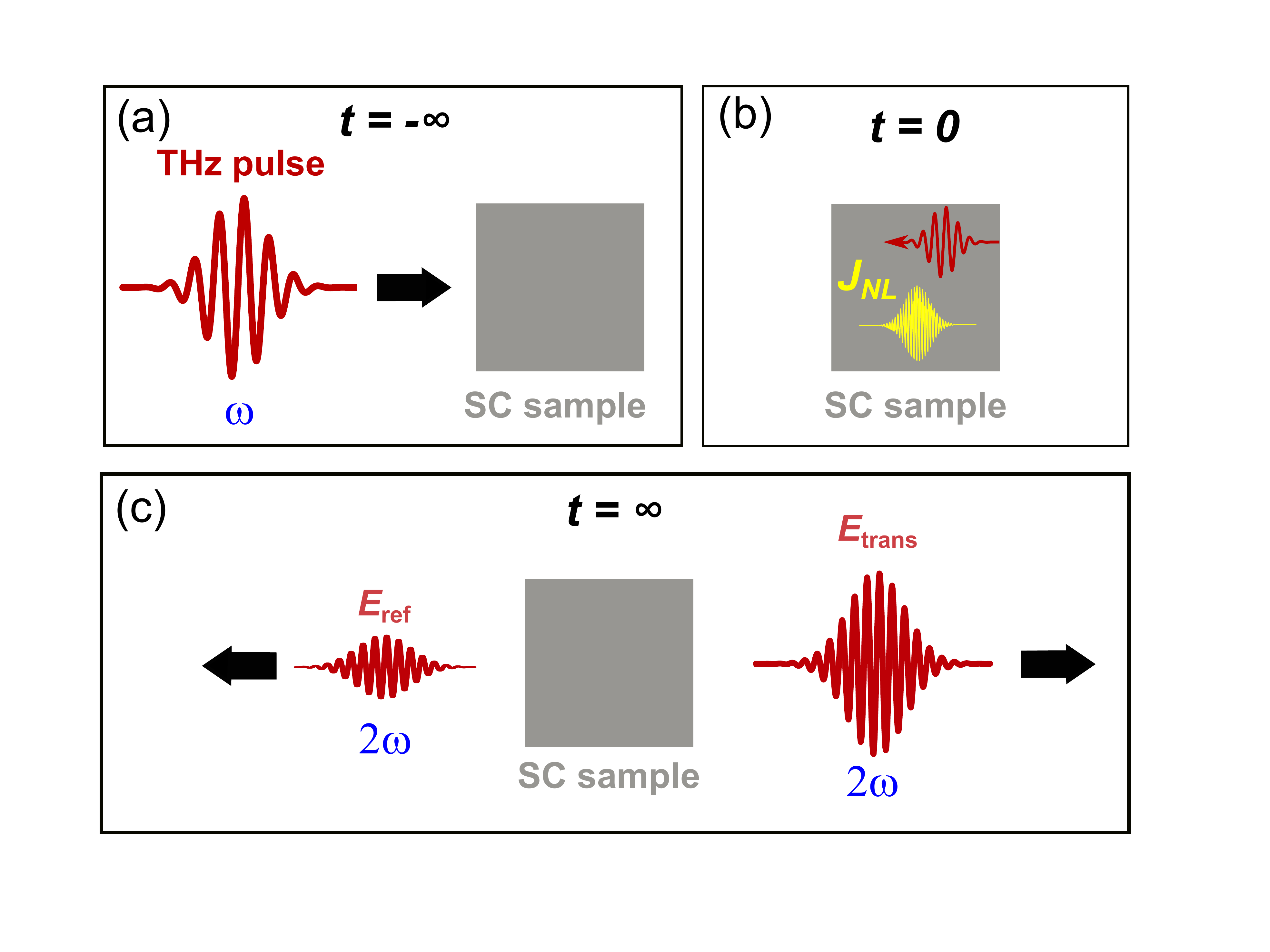}
		\caption{(Color online) Schematic representation of dc supercurrent
		 coherent nonlinear photogeneration leading to equilibrium-symmetry-forbidden second harmonic emission, non-equilibrium gapless SC, and Higgs collective modes via THz dynamical symmetry breaking .}
		\label{fig1} 
\end{figure}

Figure~\ref{fig1} illustrates 
one of the points made by  this paper: 
Nonlinear photoexcitation of the SC system together with lightwave propagation inside a superconductor thin film results in THz dynamical breaking of the equilibrium inversion symmetry, by photoinducing a 
 dc supercurrent component through nonlinear processes. Of course, the electric field from any physical source does not contain any zero-frequency dc component~\cite{Kozlov2011}, as   a consequence of Maxwell's equations:  $\int_{-\infty}^{\infty}\mathrm{d}t\,E_\mathrm{THz}(t)= 0$. However, reflected and transmitted electric fields  can show a temporal asymmetry 
and static component after interacting with a nonlinear medium~\cite{Kozlov2011}. Here, we describe such a process in superconductors
and discuss its interplay with pseudo-spin nonlinearities and 
lightwave  sub-cycle condensate acceleration. 
Our calculations demonstrate 
 nonlinear coherent  photogeneration of a dynamical broken-symmetry dc supercurrent, which  occurs via the following two steps: (1) THz excitation of the SC film with pump electric field $E_\mathrm{THz}(t)$ (red line, Fig.~\ref{fig1}(a)) induces a nonlinear ac photocurrent $J_\mathrm{NL}(t)$ (yellow line, Fig.~\ref{fig1}(b)).  (2) Analogous to four-wave mixing, this  photocurrent interferes with both forward- and reflected-propagating  THz 
 lightwaves  inside the SC, which results in coherent photogeneration of a   $\omega=0$ component via a third-order nonlinear process. The latter dc  component leads to generation of temporally asymmetric reflected ($E_\mathrm{ref}(t)$, red line Fig.~\ref{fig1}(b)) and transmitted ($E_\mathrm{trans}(t)$) electric field pulses with $\int_{-\infty}^{\infty}\mathrm{d}t\,E_{\mathrm{ref},\mathrm{trans}}(t)\neq 0$.  Such effective fields in turn drive  a dynamically induced condensate flow via  lightwave acceleration, which   breaks the equilibrium inversion symmetry of the SC system.
 This THz dynamical symmetry breaking was observed experimentally \cite{yang2019lightwave,vaswani2019discovery} 
 via  high harmonic emission at equilibrium-symmetry forbidden frequencies, particularly at the second harmonic of the THz lightwave driving frequency (red line, Fig.~\ref{fig1}(c)).
Condensate lightwave acceleration also drives  gapless SC non-equilibrium states and collective modes that last longer than the THz pulse, which 
can be controlled during cycles of  carrier wave oscillations.

This paper is organized as follows: In Sec.~\ref{sec:theory} we present the details of our gauge-invariant  density-matrix
quantum kinetic theory and the resulting gauge-invariant SC Bloch equations. The self-consistent coupling of the THz-driven SC nonlinearities to the propagating lightwave  fields is discussed in Sec.~\ref{sec:Maxwell}. In Sec.~\ref{sec:gapless} we apply our theory to demonstrate coherent nonlinear photogeneration of a  supercurrent with $\omega=0$ component, which  controls   moving condensate non-equilibrium states
with highly nonlinear responses. The experimental detection of such dynamical broken-symmetry supercurrent photogeneration 
via  high-harmonic generation is demonstrated in Sec.~\ref{sec:HHG}. The selective excitation and sensing of different classes of collective  modes of the SC order parameter, 
coherently controlled  by intense  THz  multi-cycle lightwaves,  is presented in Sec.~\ref{sec:Higgs}. We end with  conclusions.

\section{Gauge-invariant non-adiabatic theory of non-equilibrium superconductivity}
\label{sec:theory}

In this section, we derive  the gauge-invariant density matrix theory  that describes the non-equilibrium SC states driven by  lightwaves.  Here we consider  the spatially-dependent Boguliobov--de Gennes Hamiltonian for $s$-wave superconductors~\cite{Stephen1965}:
	\begin{align}
	\label{eq:Ham}
		&H=\sum_{\alpha}\int\mathrm{d}^3\mathbf{x}\,\psi_{\alpha}^\dagger(\mathbf{x})\left[\xi(\mathbf{p}+e\mathbf{A}(\mathbf{x},t))-\mu-e\phi(\mathbf{x},t) \right. \nonumber \\
		&\left.\qquad\qquad\qquad\qquad\quad
		+\mu_\mathrm{H}(\mathbf{x})+\mu^{\alpha}_\mathrm{F}(\mathbf{x})\right]\psi_{\alpha}(\mathbf{x}) \nonumber \\
		&\quad -\int\mathrm{d}^3\mathbf{x}\left[\Delta(\mathbf{x})\psi^\dagger_{\uparrow}(\mathbf{x})\psi^\dagger_{\downarrow}(\mathbf{x})+\mathrm{h.c.}\right]\,,
	\end{align}
	where the Fermionic field operators $\psi_\sigma^\dagger(\mathbf{x})$ and $\psi_\sigma(\mathbf{x})$ create and annihilate an electron with spin index $\sigma$
and the electromagnetic field is described 	by the vector potential $\mathbf{A}(\mathbf{x},t)$ and the scalar potential  $\phi(\mathbf{x},t)$. 
$\xi(\mathbf{p}+e\mathbf{A}(\mathbf{x},t))$ is the band dispersion, with momentum operator $\mathbf{p}=-\mathrm{i}\nabla_\mathbf{x}$ and electron charge $-e$ ($\hbar$=1). $\mu$ is  the chemical potential. The SC order parameter is defined as 
	\begin{align}
		\Delta(\mathbf{x})=-2\,g\langle \psi_{\downarrow}(\mathbf{x})\psi_{\uparrow}(\mathbf{x})\rangle=|\Delta(\mathbf{x})|\mathrm{e}^{\mathrm{i}\theta(\mathbf{x})}\,,
		\label{eq:gap_eq}
	\end{align}
	while 
	\begin{align}
		&\mu_\mathrm{H}(\mathbf{x})=2\sum_\sigma\int\mathrm{d}^3\mathbf{x}'\,V(\mathbf{x}-\mathbf{x}')n_{\sigma}(\mathbf{x}')
		\label{eq:mu_H}
	\end{align}
is  the Hartree energy and
	\begin{align}
		\mu^{\alpha}_\mathrm{F}(\mathbf{x})=-g\,n_{\alpha}(\mathbf{x})
		\label{eq:mu_F}
	\end{align}
	is the Fock energy, where 
	\begin{align}
		&n_{\sigma}(\mathbf{x})=\langle\psi^\dagger_{\sigma}(\mathbf{x})\psi_{\sigma}(\mathbf{x})\rangle
		\label{eq:n}
	\end{align}
	describes the spin-dependent electron populations. 
	Here,  $V(\mathbf{x})$ denotes the Coulomb potential, whose Fourier transformation is given by $V_\mathbf{q}=e^2/(\varepsilon_0 q^2)$, and $g$ describes  the effective electron--electron pairing interaction  in the BCS theory. The Hartree term moves the  in-gap Nambu--Goldstone mode up to the plasma frequency  due to the long-range Coulomb interaction  according to the Anderson--Higgs mechanism~\cite{Anderson}. The Fock energy $\mu^{\alpha}_\mathrm{F}(\mathbf{x})$ yields charge conservation of the SC system.

\subsection{Gauge-invariant density matrix equations of motion}

Gauge invariance of Hamiltonian~(\ref{eq:Ham}) under the general gauge transformation~\cite{Nambu} 
	\begin{align}
	\label{eq:gauge_trafo}
		\Psi(\mathbf{x}) \,\rightarrow\, \mathrm{e}^{\mathrm{i}\tau_3\Lambda(\mathbf{x})/2}\Psi(\mathbf{x})\,,
	\end{align}
	with the field operator in Nambu space $\Psi(\mathbf{x})=(\psi_\uparrow(\mathbf{x}),\psi_\downarrow^\dagger(\mathbf{x}))^T$ and the Pauli spin matrix $\tau_3=
\begin{pmatrix}
1 & 0 \\
0 & -1
\end{pmatrix}
$, 
is satisfied when vector potential, scalar potential and SC phase transform as
	\begin{align}
		&\mathbf{A}(\mathbf{x})\,\rightarrow\,\mathbf{A}(\mathbf{x})+\frac{c}{2e}\nabla\Lambda(\mathbf{x})\,,\nonumber \\
		&\phi(\mathbf{x})\,\rightarrow\,\phi(\mathbf{x})-\frac{1}{2e}\frac{\partial}{\partial t}\Lambda(\mathbf{x})\,,\nonumber \\
		&\zeta(\mathbf{x})\,\rightarrow\,\zeta(\mathbf{x})+\Lambda(\mathbf{x})\,.
	\end{align}
	 The  density matrix $\rho(\mathbf{x},\mathbf{x}')=\langle \Psi(\mathbf{x})^\dagger\Psi(\mathbf{x}')\rangle$, however,  depends on the specific choice of gauge.
To define a gauge invariant density matrix, we introduce center-of-mass and relative coordinates $\mathbf{R}=(\mathbf{x}+\mathbf{x}')/2$ and $\mathbf{r}=\mathbf{x}-\mathbf{x}'$ and introduce  a new density matrix~\cite{Stephen1965,Wu2017,haug}
\begin{align}
\label{eq:rho_trans}
	\tilde{\rho}(\mathbf{r},\mathbf{R})=&\mathrm{exp}\left[-\mathrm{i}e\int_0^\frac{1}{2}\mathrm{d}\lambda\,\mathbf{A}(\mathbf{R}+\lambda\,\mathbf{r},t)\cdot\mathbf{r}\,\tau_3\right]\,\rho(\mathbf{r},\mathbf{R}) \nonumber \\ 
	&\times\mathrm{exp}\left[-\mathrm{i}e\int^0_{-\frac{1}{2}}\mathrm{d}\lambda\,\mathbf{A}(\mathbf{R}+\lambda\,\mathbf{r},t)\cdot\mathbf{r}\,\tau_3\right]\,,
\end{align}
where $\rho(\mathbf{r},\mathbf{R})=\langle\Psi^\dagger(\mathbf{R}+\frac{\mathbf{r}}{2})\Psi(\mathbf{R}-\frac{\mathbf{r}}{2}))\rangle$ is the Wigner function. By applying the gauge transformation (\ref{eq:gauge_trafo}),  $\tilde{\rho}(\mathbf{r},\mathbf{R})$ transforms as~\cite{Wu2017}
\begin{align}
\label{rho_gt}
	\tilde{\rho}(\mathbf{r},\mathbf{R})\,\rightarrow\,\mathrm{exp}\left[\mathrm{i}\tau_3\Lambda(\mathbf{R})/2\right]\tilde{\rho}(\mathbf{r},\mathbf{R})\mathrm{exp}\left[-\mathrm{i}\tau_3\Lambda(\mathbf{R})/2\right]\,.
\end{align}
Unlike for  the transformed phase of $\rho(\mathbf{r},\mathbf{R})$, which is generally a function of both coordinates $\mathbf{R}$ and $\mathbf{r}$, the phase $\Lambda({\bf R})$ in Eq.~(\ref{rho_gt}) depends only on the center-of-mass coordinate. This allows for a  gauge-invariant density matrix description of non-equilibrium SC dynamics.

The time evolution  of the density matrix (\ref{eq:rho_trans}) is obtained by using the Heisenberg equation of motion
\begin{equation}
	\mathrm{i}\frac{\partial}{\partial t}\tilde{\rho}=\langle \left[\tilde{\rho},H\right]\rangle\,.
\end{equation}
We apply the Fourier transformation with respect to the relative coordinate $\mathbf{r}$,
\begin{equation}
	\tilde{\rho}(\mathbf{k},\mathbf{R})=\int\mathrm{d}^3\mathbf{r}\,\tilde{\rho}(\mathbf{r},\mathbf{R})\,\mathrm{e}^{-\mathrm{i}\mathbf{k}\cdot\mathbf{r}}\,.
	\label{eq:rho_FT}
\end{equation}
The equation of motion for $\tilde{\rho}(\mathbf{k},\mathbf{R})$
has  contributions of the form $\Delta(\mathbf{R}+\frac{\mathrm{i}}{2}\nabla_\mathbf{k})\tilde{\rho}(\mathbf{k},\mathbf{R})$,  which are evaluated  by applying the gradient expansion 
\begin{equation}
\label{eq:gradient}
	\Delta(\mathbf{R}+\frac{\mathrm{i}}{2}\nabla_\mathbf{k})=\sum_{n=0}^\infty\left(\frac{\mathrm{i}}{2}\right)^n\frac{(\nabla_\mathbf{R}\cdot\nabla_\mathbf{k})^n}{n!}\Delta(\mathbf{R})\,.
\end{equation}
Similar to Ginzburg--Landau theory, this expansion in powers of $\nabla_\mathbf{R}\cdot\nabla_\mathbf{k}$ 
can be truncated when the characteristic length for spatial variation of the SC condensate (center of mass) exceeds the coherence length of the Cooper pair 
(relative motion). 
To simplify the equations of motion, we also apply the gauge transformation
\begin{align}
		\tilde{\rho}(\mathbf{k},\mathbf{R})=\mathrm{e}^{-\mathrm{i}\tau_3\theta(\mathbf{R})/2}\tilde{\rho}(\mathbf{k},\mathbf{R})\mathrm{e}^{\mathrm{i}\tau_3\theta(\mathbf{R})/2}\,,
		\label{eq:phase-trafo}
\end{align}
which eliminates the phase of the SC order parameter in the equations of motion. After applying the above gradient expansion and the unitary transformation (\ref{eq:phase-trafo}), we obtain the following exact gauge-invariant spatially-dependent SC Bloch equations:
\begin{widetext}
\begin{align}
\label{eq:eom-full1}
	\mathrm{i}\frac{\partial}{\partial t}\tilde{\rho}_{1,1}(\mathbf{k},\mathbf{R})&=\left[\xi\left(-\mathbf{k}-\frac{\mathrm{i}}{2}\nabla_\mathbf{R}+\mathrm{i}\frac{e}{2}\sum_{n=0}^\infty\left(-\frac{1}{4}\right)^n\frac{(\nabla_\mathbf{k}\cdot\nabla_\mathbf{R})^{2n}}{(2n+1)!}\nabla_\mathbf{k}\times\mathbf{B}(\mathbf{R})\right.\right. \nonumber \\
	&\left.\left.\qquad\quad\;-e\sum_{n=1}^\infty 2n\left(-\frac{1}{4}\right)^n\frac{(\nabla_\mathbf{k}\cdot\nabla_\mathbf{R})^{2n-1}}{(2n+1)!}\nabla_\mathbf{k}\times\mathbf{B}(\mathbf{R})\right)\right. \nonumber \\ 
	&\quad\;\;\left.-\xi\left(-\mathbf{k}+\frac{\mathrm{i}}{2}\nabla_\mathbf{R}-\mathrm{i}\frac{e}{2}\sum_{n=0}^\infty\left(-\frac{1}{4}\right)^n\frac{(\nabla_\mathbf{k}\cdot\nabla_\mathbf{R})^{2n}}{(2n+1)!}\nabla_\mathbf{k}\times\mathbf{B}(\mathbf{R})\right.\right. \nonumber \\
	&\left.\left.\qquad\quad\;-e\sum_{n=1}^\infty 2n\left(-\frac{1}{4}\right)^n\frac{(\nabla_\mathbf{k}\cdot\nabla_\mathbf{R})^{2n-1}}{(2n+1)!}\nabla_\mathbf{k}\times\mathbf{B}(\mathbf{R})\right)\right. \nonumber \\ 
	&\quad\;\;\left. -2\sum_{n=0}^\infty\frac{\left(\frac{\mathrm{i}}{2}\right)^{2n+1}(\nabla_\mathbf{k}\cdot\nabla_\mathbf{R})^{2n+1}}{(2n+1)!}\left(\mu_\mathrm{H}(\mathbf{R})+\mu_\mathrm{F}^{\uparrow}(\mathbf{R})\right)
	\right. \nonumber \\ 
	&\quad\;\;\left.-\mathrm{i}e\sum_{n=0}^\infty\frac{(\nabla_\mathbf{k}\cdot\nabla_\mathbf{R})^{2n}}{(2n+1)!}\left(-\frac{1}{4}\right)^n\,\mathbf{E}(\mathbf{R})\cdot\nabla_\mathbf{k}\right]\tilde{\rho}_{1,1}(\mathbf{k},\mathbf{R}) \nonumber \\
	&+\mathrm{exp}\left[\frac{\mathrm{i}}{2}\nabla_\mathbf{R}\cdot\nabla_\mathbf{k}\right]|\Delta(\mathbf{R})|\;\mathrm{exp}\left[-\frac{1}{2}\sum_{n=0}^\infty\frac{(\nabla_\mathbf{k}\cdot\nabla_\mathbf{R})^n}{(n+1)!}\left(\frac{\mathrm{i}}{2}\right)^n\,\mathbf{p}_\mathrm{S}(\mathbf{R})\cdot\nabla_\mathbf{k}\right]\tilde{\rho}_{2,1}(\mathbf{k},\mathbf{R})\nonumber \\
	&-\mathrm{exp}\left[-\frac{\mathrm{i}}{2}\nabla_\mathbf{R}\cdot\nabla_\mathbf{k}\right]|\Delta(\mathbf{R})|\;\mathrm{exp}\left[-\frac{1}{2}\sum_{n=0}^\infty\frac{(\nabla_\mathbf{k}\cdot\nabla_\mathbf{R})^n}{(n+1)!}\left(-\frac{\mathrm{i}}{2}\right)^n\,\mathbf{p}_\mathrm{S}(\mathbf{R})\cdot\nabla_\mathbf{k}\right]\tilde{\rho}_{1,2}(\mathbf{k},\mathbf{R})\,,
\end{align}
\begin{align}	
\label{eq:eom-full2}
	\mathrm{i}\frac{\partial}{\partial t}\tilde{\rho}_{2,2}(\mathbf{k},\mathbf{R})&=\left[\xi\left(\mathbf{k}-\frac{\mathrm{i}}{2}\nabla_\mathbf{R}-\mathrm{i}\frac{e}{2}\sum_{n=0}^\infty\left(-\frac{1}{4}\right)^n\frac{(\nabla_\mathbf{k}\cdot\nabla_\mathbf{R})^{2n}}{(2n+1)!}\nabla_\mathbf{k}\times\mathbf{B}(\mathbf{R})\right.\right. \nonumber \\
	&\left.\left.\qquad\quad\;+e\sum_{n=1}^\infty 2n\left(-\frac{1}{4}\right)^n\frac{(\nabla_\mathbf{k}\cdot\nabla_\mathbf{R})^{2n-1}}{(2n+1)!}\nabla_\mathbf{k}\times\mathbf{B}(\mathbf{R})\right)\right. \nonumber \\ 
	&\quad\;\;\left.-\xi\left(\mathbf{k}+\frac{\mathrm{i}}{2}\nabla_\mathbf{R}+\mathrm{i}\frac{e}{2}\sum_{n=0}^\infty\left(-\frac{1}{4}\right)^n\frac{(\nabla_\mathbf{k}\cdot\nabla_\mathbf{R})^{2n}}{(2n+1)!}\nabla_\mathbf{k}\times\mathbf{B}(\mathbf{R})\right.\right. \nonumber \\
	&\left.\left.\qquad\quad\;+e\sum_{n=1}^\infty 2n\left(-\frac{1}{4}\right)^n\frac{(\nabla_\mathbf{k}\cdot\nabla_\mathbf{R})^{2n-1}}{(2n+1)!}\nabla_\mathbf{k}\times\mathbf{B}(\mathbf{R})\right)\right. \nonumber \\ 
	&\quad\;\;\left. +2\sum_{n=0}^\infty\frac{\left(\frac{\mathrm{i}}{2}\right)^{2n+1}(\nabla_\mathbf{k}\cdot\nabla_\mathbf{R})^{2n+1}}{(2n+1)!}\left(\mu_\mathrm{H}(\mathbf{R})+\mu_\mathrm{F}^{\downarrow}(\mathbf{R})\right)\right. \nonumber \\	 
	&\quad\;\;\left.+\mathrm{i}e\sum_{n=0}^\infty\frac{(\nabla_\mathbf{k}\cdot\nabla_\mathbf{R})^{2n}}{(2n+1)!}\left(-\frac{1}{4}\right)^n\,\mathbf{E}(\mathbf{R})\cdot\nabla_\mathbf{k}\right]\tilde{\rho}_{2,2}(\mathbf{k},\mathbf{R}) \nonumber \\
	&-\mathrm{exp}\left[-\frac{\mathrm{i}}{2}\nabla_\mathbf{R}\cdot\nabla_\mathbf{k}\right]|\Delta(\mathbf{R})|\;\mathrm{exp}\left[\frac{1}{2}\sum_{n=0}^\infty\frac{(\nabla_\mathbf{k}\cdot\nabla_\mathbf{R})^n}{(n+1)!}\left(-\frac{\mathrm{i}}{2}\right)^n\,\mathbf{p}_\mathrm{S}(\mathbf{R})\cdot\nabla_\mathbf{k}\right]\tilde{\rho}_{2,1}(\mathbf{k},\mathbf{R})\nonumber \\
	&+\mathrm{exp}\left[\frac{\mathrm{i}}{2}\nabla_\mathbf{R}\cdot\nabla_\mathbf{k}\right]|\Delta(\mathbf{R})|\;\mathrm{exp}\left[\frac{1}{2}\sum_{n=0}^\infty\frac{(\nabla_\mathbf{k}\cdot\nabla_\mathbf{R})^n}{(n+1)!}\left(\frac{\mathrm{i}}{2}\right)^n\,\mathbf{p}_\mathrm{S}(\mathbf{R})\cdot\nabla_\mathbf{k}\right]\tilde{\rho}_{1,2}(\mathbf{k},\mathbf{R})\,,
	\end{align}
\begin{align}
\label{eq:eom-full3}
	\mathrm{i}\frac{\partial}{\partial t}\tilde{\rho}_{1,2}(\mathbf{k},\mathbf{R})&=\left[-\xi\left(-\mathbf{k}+\frac{\mathrm{i}}{2}\nabla_\mathbf{R}-e\sum_{n=0}^\infty(2n+1)\left(\frac{\mathrm{i}}{2}\right)^{2n+1}\frac{(\nabla_\mathbf{k}\cdot\nabla_\mathbf{R})^{2n}}{(2n+2)!}\nabla_\mathbf{k}\times\mathbf{B}(\mathbf{R})\right.\right. \nonumber \\
		&\qquad\qquad\left.\left.-\mathrm{i}\frac{e}{2}\sum_{n=0}^\infty\left(\frac{\mathrm{i}}{2}\right)^{2n+1}\frac{(\nabla_\mathbf{k}\cdot\nabla_\mathbf{R})^{2n+1}}{(2n+2)!}\nabla_\mathbf{k}\times\mathbf{B}(\mathbf{R})-\mathbf{p}_\mathrm{S}^{\nu_0}(\mathbf{R})/2\right)\right. \nonumber \\ 
	&\quad\;\;\left.-\xi\left(\mathbf{k}+\frac{\mathrm{i}}{2}\nabla_\mathbf{R}+e\sum_{n=0}^\infty(2n+1)\left(\frac{\mathrm{i}}{2}\right)^{2n+1}\frac{(\nabla_\mathbf{k}\cdot\nabla_\mathbf{R})^{2n}}{(2n+2)!}\nabla_\mathbf{k}\times\mathbf{B}(\mathbf{R})\right.\right. \nonumber \\
		&\qquad\qquad\left.\left.+\mathrm{i}\frac{e}{2}\sum_{n=0}^\infty\left(\frac{\mathrm{i}}{2}\right)^{2n+1}\frac{(\nabla_\mathbf{k}\cdot\nabla_\mathbf{R})^{2n+1}}{(2n+2)!}\nabla_\mathbf{k}\times\mathbf{B}(\mathbf{R})-\mathbf{p}_\mathrm{S}^{\nu_0}(\mathbf{R})/2\right)-2\mu_\mathrm{eff}(\mathbf{R})\right. \nonumber \\
	&\quad\;\;\left. -2\sum_{n=0}^\infty\frac{\left(\frac{\mathrm{i}}{2}\right)^{2n}(\nabla_\mathbf{k}\cdot\nabla_\mathbf{R})^{2n}}{(2n)!}\mu_\mathrm{H}(\mathbf{R})-\sum_{n=0}^\infty\frac{\left(-\frac{\mathrm{i}}{2}\right)^n(\nabla_\mathbf{k}\cdot\nabla_\mathbf{R})^n}{n!}\left(\mu_\mathrm{F}^{\downarrow}(\mathbf{R})+(-1)^n\mu_\mathrm{F}^{\uparrow}(\mathbf{R})\right) \right. \nonumber \\
	&\quad\;\;\left. -\mathrm{i}\,e\,\sum_{n=0}^\infty\frac{(\nabla_\mathbf{k}\cdot\nabla_\mathbf{R})^{2n+1}}{(2n+2)!}\left(\frac{\mathrm{i}}{2}\right)^{2n+1}\,\mathbf{E}(\mathbf{R})\cdot\nabla_\mathbf{k}\right]\tilde{\rho}_{1,2}(\mathbf{k},\mathbf{R})\nonumber \\
	&+\mathrm{exp}\left[\frac{\mathrm{i}}{2}\nabla_\mathbf{R}\cdot\nabla_\mathbf{k}\right]|\Delta(\mathbf{R})|\mathrm{exp}\left[-\frac{1}{2}\sum_{n=0}^\infty\frac{(\nabla_\mathbf{k}\cdot\nabla_\mathbf{R})^n}{(n+1)!}\left(\frac{\mathrm{i}}{2}\right)^n\,\mathbf{p}_\mathrm{S}(\mathbf{R})\cdot\nabla_\mathbf{k}\right]\tilde{\rho}_{2,2}(\mathbf{k},\mathbf{R})\nonumber \\
	&-\mathrm{exp}\left[-\frac{\mathrm{i}}{2}\nabla_\mathbf{R}\cdot\nabla_\mathbf{k}\right]|\Delta(\mathbf{R})|\mathrm{exp}\left[\frac{1}{2}\sum_{n=0}^\infty\frac{(\nabla_\mathbf{k}\cdot\nabla_\mathbf{R})^n}{(n+1)!}\left(-\frac{\mathrm{i}}{2}\right)^n\,\mathbf{p}_\mathrm{S}(\mathbf{R})\cdot\nabla_\mathbf{k}\right]\tilde{\rho}_{1,1}(\mathbf{k},\mathbf{R})\,.
\end{align}
\end{widetext}
Here we introduced the gauge-invariant superfluid momentum and effective chemical potential
\begin{align}
		&\mathbf{p}_\mathrm{S}(\mathbf{R},t)=\nabla_\mathbf{R}\theta(\mathbf{R},t)-2e\mathbf{A}(\mathbf{R},t)\,,\nonumber \\  &\mu_\mathrm{eff}(\mathbf{R},t)=e\,\phi(\mathbf{R},t)+\frac{1}{2}\frac{\partial}{\partial t}\theta(\mathbf{R},t)-\mu\,, \label{mueff} 
\end{align}
and identified the electric and magnetic fields 
\begin{align}
	\mathbf{E}(\mathbf{R})=-\nabla_\mathbf{R}\phi(\mathbf{R})-\frac{\partial}{\partial t}\mathbf{A}(\mathbf{R})\,,\quad	  \mathbf{B}(\mathbf{R})=\nabla_\mathbf{R}\times\mathbf{A}(\mathbf{R})\,.
\end{align}
The Higgs collective mode corresponds to amplitude fluctuations of the SC order parameter
\begin{widetext}
\begin{align}
\Delta(\mathbf{R})=-2g\sum_\mathbf{k}\mathrm{exp}\left[e\sum_{n=0}^\infty\left(\frac{\mathrm{i}}{2}\right)^{2n+1}\frac{(\nabla_\mathbf{k}\cdot\nabla_\mathbf{R})^{2n+1}}{(2n+2)!}\mathbf{p}_\mathrm{S}(\mathbf{R})\cdot\nabla_\mathbf{k}\right]\tilde{\rho}_{2,1}(\mathbf{k},\mathbf{R})\,.
\end{align}
\end{widetext}
The light-driven  dynamics of the phase of the SC order parameter $\Delta(\mathbf{R})$ is determined by the equation of motion
\begin{widetext}
\begin{align}
\frac{\partial}{\partial t}\theta(\mathbf{R})&=-2\left(e\,\phi(\mathbf{R})-\,\mu-\mu_\mathrm{H}(\mathbf{R})-\frac{1}{2}(\mu_\mathrm{F}^{\uparrow}(\mathbf{R})+\mu_\mathrm{F}^{\downarrow}(\mathbf{R}))\right)\nonumber \\
&+\frac{g}{|\Delta(\mathbf{R})|}\sum_\mathbf{k} \left(\left[\xi(\mathbf{k}+\frac{\mathrm{i}}{2}\nabla_\mathbf{R}-\mathbf{p}_\mathrm{S}(\mathbf{R})/2)+\xi(\mathbf{k}-\frac{\mathrm{i}}{2}\nabla_\mathbf{R}+\mathbf{p}_\mathrm{S}(\mathbf{R})/2)\right]\tilde{\rho}_{1,2}(\mathbf{k},\mathbf{R})\right. \nonumber \\ 
&\left.\qquad\qquad\qquad\qquad\quad+\left[\xi(\mathbf{k}-\frac{\mathrm{i}}{2}\nabla_\mathbf{R}-\mathbf{p}_\mathrm{S}(\mathbf{R})/2)+\xi(\mathbf{k}+\frac{\mathrm{i}}{2}\nabla_\mathbf{R}+\mathbf{p}_\mathrm{S}(\mathbf{R})/2)\right]\tilde{\rho}_{2,1}(\mathbf{k},\mathbf{R})\right)\nonumber \\
&+\frac{2g}{|\Delta(\mathbf{R})|}\sum_\mathbf{k} |\Delta(\mathbf{R})|\left[\tilde{\rho}_{1,1}(\mathbf{k},\mathbf{R})-\tilde{\rho}_{2,2}(\mathbf{k},\mathbf{R})\right]\,.
\end{align}
\end{widetext}
The lightwave  field  accelerates the  center-of-mass  of the Cooper-pairs  with gauge-invariant momentum determined by the electric field and by spatial fluctuations: 
\begin{align}
\frac{\partial}{\partial t}\mathbf{p}_\mathrm{S}(\mathbf{R},t)&=2 \nabla_\mathbf{R} \mu_{\rm{eff}}({\bf R},t)
+2e\,\mathbf{E}(\mathbf{R},t).
\, \label{ps} 
\end{align}
As seen from 
Eq.~(\ref{mueff}), 
the time-dependent changes in the 
SC order parameter phase are included in  the above equation of motion
and   
determine the condensate 
center-of-mass momentum 
$\mathbf{p}_\mathrm{S}({\bf R},t)$.
The latter develops here 
as a result 
 of  lightwave acceleration of the macroscopic Cooper pair state.
 This acceleration is strong 
 in the case of intense THz fields
available today and    
therefore we include 
$\mathbf{p}_\mathrm{S}({\bf R},t)$
in the above density matrix equations of motion 
without  perturbative expansions.
For example,  lightwave acceleration displaces the  populations and coherences in momentum space by $\mathbf{p}_\mathrm{S}(t)/2$, which 
unlike in previous works  are treated exactly here. 
In this way, 
we describe the breaking of the equilibrium inversion symmetry of electron ($\tilde{\rho}_{1,1}(\mathbf{k},\mathbf{R})$) and hole  ($\tilde{\rho}_{2,2}(\mathbf{k},\mathbf{R})$) populations, 
as the  condensate momentum vector defines a preferred direction.
The lightwave condensate  acceleration is described by quantum transport terms of the form $\mathbf{E}(\mathbf{R})\cdot\nabla_\mathbf{k}\tilde{\rho}(\mathbf{k},\mathbf{R})$ in the equations of motion (\ref{eq:eom-full1})--(\ref{eq:eom-full3}). The latter are absent in the pseudo-spin precession model~\cite{Forster2017}
and lead to linear couplings of the electric field.  
Higher orders in the spatial gradient expansion, e.~g.~the  first ($\mathcal{O}(\nabla_\mathbf{k}\cdot\nabla_\mathbf{R})$) and second order terms ($\mathcal{O}((\nabla_\mathbf{k}\cdot\nabla_\mathbf{R})^2)$), contain the kinetic terms in the Ginzburg--Landau equation~\cite{Wu:2018}.
Such spatial contributions can be expanded as in  Ginzburg--Landau theory.

The Fock energies in the above gauge-invariant density matrix equations of motion,
\begin{widetext}
\begin{align}
\label{eq:mu_f}
\mu_\mathrm{F}^{\uparrow}(\mathbf{R})=-g\sum_\mathbf{k}&\mathrm{exp}\left[e\sum_{n=0}^\infty\left(\frac{\mathrm{i}}{2}\right)^{2n}\frac{(\nabla_\mathbf{k}\cdot\nabla_\mathbf{R})^{2n}}{(2n+1)!}\mathbf{p}_\mathrm{S}(\mathbf{R})\cdot\nabla_\mathbf{k}\right]\tilde{\rho}_{1,1}(\mathbf{k},\mathbf{R})\,, \nonumber \\
\mu_\mathrm{F}^{\downarrow}(\mathbf{R})=-g\sum_\mathbf{k}&\left(1-\mathrm{exp}\left[-e\sum_{n=0}^\infty\left(\frac{\mathrm{i}}{2}\right)^{2n}\frac{(\nabla_\mathbf{k}\cdot\nabla_\mathbf{R})^{2n}}{(2n+1)!}\mathbf{p}_\mathrm{S}(\mathbf{R})\cdot\nabla_\mathbf{k}\right]\tilde{\rho}_{2,2}(\mathbf{k},\mathbf{R})\right)\,,
\end{align}
\end{widetext}
ensure charge conservation of the SC system. In particular, the gauge-invariant current,
\begin{align}
\mathbf{J}(\mathbf{R})=e\sum_\mathbf{k}\nabla_\mathbf{k}\xi(\mathbf{k})\left[\tilde{\rho}_{1,1}(\mathbf{k},\mathbf{R})+\tilde{\rho}_{2,2}(\mathbf{k},\mathbf{R})\right]\,, 
\label{eq:current} 
\end{align}
and electron density,
\begin{align}
n(\mathbf{R})=\sum_\mathbf{k}\left[1+\tilde{\rho}_{1,1}(\mathbf{k},\mathbf{R})-\tilde{\rho}_{2,2}(\mathbf{k},\mathbf{R})\right]\,,
\end{align}
explicitly satisfy the continuity equation
\begin{align}
	e\frac{\partial}{\partial t}n(\mathbf{R})+\nabla_\mathbf{R}\cdot \mathbf{J}(\mathbf{R})=0\,,
\end{align}
which is a direct consequence of the gauge invariance of the equations of motion (\ref{eq:eom-full1})--(\ref{eq:eom-full3}).
The  SC phase and amplitude 
dynamics as well as spatial dependence  are thus  treated consistently.

\subsection{Homogeneous SC system}

For  weak spatial dependence  and  homogeneous excitation conditions, we neglect all terms of order $\mathcal{O}(\nabla_\mathbf{k}\cdot\nabla_\mathbf{R})$ and higher in the gradient expansion (\ref{eq:gradient}), as well as the $\mathbf{R}$-dependence of $\mathbf{E}$- and $\mathbf{B}$-fields,  in the equations of motion (\ref{eq:eom-full1})--(\ref{eq:eom-full3}). We then  obtain the gauge-invariant homogenous SC Bloch equations valid  for any 
 condensate center-of-mass momentum ${\bf p}_\mathrm{S}(t)$:
\begin{widetext}
\begin{align}
\label{eq:eoms}
	&\mathrm{i}\frac{\partial}{\partial t}\tilde{\rho}_{1,1}(\mathbf{k})=-\mathrm{i}\,e\,\mathbf{E}(t)\cdot\nabla_\mathbf{k}\tilde{\rho}_{1,1}(\mathbf{k})-|\Delta|\left[\tilde{\rho}_{1,2}(\mathbf{k}-\mathbf{p}_\mathrm{S}/2)-\tilde{\rho}_{2,1}(\mathbf{k}-\mathbf{p}_\mathrm{S}/2)\right]\,, \nonumber \\
	&\mathrm{i}\frac{\partial}{\partial t}\tilde{\rho}_{2,2}(\mathbf{k})=\mathrm{i}\,e\,\mathbf{E}(t)\cdot\nabla_\mathbf{k}\tilde{\rho}_{2,2}(\mathbf{k})+|\Delta|\left[\tilde{\rho}_{1,2}(\mathbf{k}+\mathbf{p}_\mathrm{S}/2)-\tilde{\rho}_{2,1}(\mathbf{k}+\mathbf{p}_\mathrm{S}/2)\right]\,, \nonumber \\ 
	&\mathrm{i}\frac{\partial}{\partial t}\tilde{\rho}_{1,2}(\mathbf{k})=-[\xi(\mathbf{k}-\mathbf{p}_\mathrm{S}/2)+\xi(-\mathbf{k}-\mathbf{p}_\mathrm{S}/2)+2(\mu_\mathrm{eff}+\mu_\mathrm{F})]\tilde{\rho}_{1,2}(\mathbf{k})\nonumber \\
	&\qquad\qquad\quad\;\;+|\Delta|\left[\tilde{\rho}_{2,2}(\mathbf{k}-\mathbf{p}_\mathrm{S}/2)-\tilde{\rho}_{1,1}(\mathbf{k}+\mathbf{p}_\mathrm{S}/2)\right]\,,
\end{align}
\end{widetext}
where 
\begin{align}
\label{eq:SC-homo}
&\mathbf{p}_\mathrm{S}=-2\,e\,\mathbf{A}\,,\quad \mu_\mathrm{eff}=e\,\phi+\frac{1}{2}\frac{\partial}{\partial t}\theta-\mu\,,\nonumber \\ &|\Delta|=-2g\sum_\mathbf{k} \tilde{\rho}_{2,1}(\mathbf{k})\,,\nonumber \\ &\mu_\mathrm{F}\equiv\frac{1}{2}\left(\mu^{\downarrow}_\mathrm{F}+\mu^{\uparrow}_\mathrm{F}\right)=-g\sum_\mathbf{k}\left[1+\tilde{\rho}_{1,1}(\mathbf{k})-\tilde{\rho}_{2,2}(\mathbf{k})\right]\,.
\end{align}
The equations of motion for condensate momentum $\mathbf{p}_\mathrm{S}$ and SC order parameter phase $\theta$ simplify to
\begin{align}
\frac{\partial}{\partial t}\mathbf{p}_\mathrm{S}=2e\,\mathbf{E}\,
\end{align}
and 
\begin{align}
&\frac{\partial}{\partial t}\theta=-2\left(e\,\phi-\,\mu-\mu_\mathrm{F}\right)\nonumber \\
&+\frac{g}{|\Delta|}\sum_\mathbf{k}\left[\xi(\mathbf{k}-\mathbf{p}_\mathrm{S}/2)+\xi(\mathbf{k}+\mathbf{p}_\mathrm{S}/2)\right]\left(\tilde{\rho}_{1,2}(\mathbf{k})+\tilde{\rho}_{2,1}(\mathbf{k})\right)\nonumber \\
&+\frac{2g}{|\Delta|}\sum_\mathbf{k} |\Delta|\left[\tilde{\rho}_{1,1}(\mathbf{k})-\tilde{\rho}_{2,2}(\mathbf{k})\right]. 
\end{align}
The   gauge-invariant Bloch equations  Eq.~(\ref{eq:eoms}) reduce to the Anderson pseudo-spin precession model by omitting the  transport terms $\propto \mathbf{E}(t)$, the ${\bf p}_\mathrm{S}(t)/2$-displacement of populations and coherences, and the SC phase and Fock contributions to the  chemical potential.

There are three ways in which the  lightwave  fields couple to the SC in Eq.~(\ref{eq:eoms}) in the spatially homogeneous limit:
(1) First, the familiar minimal coupling, $ \xi(\mathbf{k}-\mathbf{p}_\mathrm{S}/2)+\xi(-\mathbf{k}-\mathbf{p}_\mathrm{S}/2)$, drives  even-order nonlinearities 
of the SC order parameter.
This coupling depends on the electron band dispersion 
non-parabolicity and can be expanded 
in  $\mathcal{O}(p_s^{2n})$=$\mathcal{O}(A^{2n})$  even  terms 
 ~\cite{matsunaga2014}. 
 This coupling does not contribute to the linear response \cite{Pekker2015} and has been  
 studied before in the context of the Anderson pseudo-spin precession model. 
(2)  Second, the   condensate acceleration by the lightwave effective field 
results in SC order parameter nonlinearities that 
are of odd order in the electric field.  These THz nonlinear quantum transport contributions  come from terms of the form  $\mathrm{i}\,e\mathbf{E}(t)\cdot\nabla_\mathbf{k}\tilde{\rho}(\mathbf{k})$ in Eq.~(\ref{eq:eoms}).
(3)  Third, for accelerated condensate
with finite center-of-mass momentum,
 the 
population and coherence  displacements in momentum space 
by $ \pm \mathbf{p}_\mathrm{S}/2$  are treated non-perturbatively for intense THz fields.  
In particular, 
with lightwave acceleration, 
we thus describe a non-equilibrium moving condensate consisting of 
 Cooper pairs  formed by $({\bf k} + \mathbf{p}_\mathrm{S}(t)/2, \uparrow)$
and $(-{\bf k} + \mathbf{p}_\mathrm{S}(t)/2, \downarrow)$ electrons. 
A large momentum ${\bf p}_\mathrm{S}(t)$ then leads to a  lightwave-induced anisotropy 
in momentum space, which  results in   new non-perturbative 
contributions for $\Delta \ne 0$.
Note that ${\bf p}_\mathrm{S}(t)$ is determined by 
self-consistent non-perturbative coupling between the propagating electromagnetic 
fields and the nonlinear supercurrent, discussed next.  
	
\section{Self-consistent coupling between propagating electromagnetic  field and nonlinear supercurrent}
\label{sec:Maxwell}

To include lightwave propagation effects, we use  from Maxwell's wave equation for the electric field~\cite{zhou1991,Hirsch} 
\begin{align}
\left[\nabla_\mathbf{r}^2-\frac{n(\mathbf{r})^2}{c^2}\frac{\partial^2}{\partial t^2}\right]\mathbf{E}(\mathbf{r},t)=-\mu_0\frac{\partial}{\partial t}\mathbf{J}(\mathbf{r},t)\,
\end{align}
for  background refractive index $n(\mathbf{r})$. 
The above equation describes the  effective 
lightwave field that drives the SC condensate, which is modified 
as compared to the applied laser field due to the 
coupling  with the nonlinear supercurrent 
Eq.~(\ref{eq:current}).
By decomposing the electric field into components parallel and perpendicular ($z$-direction) to the SC film, $\mathbf{E}=\mathbf{E}_\perp+\mathbf{E}_\parallel$, and applying Fourier transformation with respect to the in-plane coordinates $(x,y)={\bf \rho}$, 
\begin{align}
\mathbf{E}(\rho,z,t)=\frac{1}{S}\sum_{\mathbf{q}_\parallel}\mathrm{e}^{\mathrm{i}\mathbf{q}_\parallel\cdot\rho}\mathbf{E}(\mathbf{q}_\parallel,z,t)\,,
\end{align}
we transform  the wave equation for the in-plane electric field into 
\begin{align}
\left[\frac{\partial^2}{\partial z^2}-q_\parallel^2-\frac{n(z)^2}{c^2}\frac{\partial^2}{\partial t^2}\right]\mathbf{E}_\parallel(\mathbf{q}_\parallel,z,t)=-\mu_0\frac{\partial}{\partial t}\mathbf{J}_\parallel(\mathbf{q}_\parallel,z,t)\,.
\end{align}
We next assume that the externally applied electric field propagates perpendicular to the film, such that the wave vector parallel to the film vanishes, i.~e. $\mathbf{q}_\parallel=0$, and there is only $z$-dependence.  As a result, lightwave propagation inside the SC film can be described by the one-dimensional wave equation 
\begin{align}
\left[\frac{\partial^2}{\partial z^2}-\frac{n(z)^2}{c^2}\frac{\partial^2}{\partial t^2}\right]\mathbf{E}_\parallel(z,t)=-\mu_0\frac{\partial}{\partial t}\mathbf{J}_\parallel(z,t)\,.
\end{align}
When the wavelength of the applied laser  field exceeds the thickness of the SC film, we can approximate the $z$-dependence of the current by $\mathbf{J}_\parallel(z,t)=\delta(z)\mathbf{J}_\parallel(t)$, such that the wave equation becomes
\begin{align}
\label{eq:wave-eq_final}
\left[\frac{\partial^2}{\partial z^2}-\frac{n(z)^2}{c^2}\frac{\partial^2}{\partial t^2}\right]\mathbf{E}_\parallel(z,t)=-\mu_0\frac{\partial}{\partial t}\delta(z)\mathbf{J}_\parallel(t)\,.
\end{align}
Assuming that substrate and SC film have comparable background refractive index $n$, Eq.~(\ref{eq:wave-eq_final}) can be solved analytically, yielding the self-consistent electric field~\cite{Jahnke1997}
\begin{align}
\label{eq:wave_eq-sol}
\mathbf{E}(z,t)=\mathbf{E}_0(z,t)-\mu_0\frac{c}{2n}\mathbf{J}_\parallel(t-|nz|/c)\,,
\end{align}
where $\mathbf{E}_0(z,t)$ denotes the externally applied electric field, incident on the SC film from the left, i.~e. $z<0$. The reflected and transmitted electric fields are given by ~\cite{Jahnke1997} 
\begin{align}
&\mathbf{E}_\mathrm{ref}(z,t)=-\mu_0\frac{c}{2n}\mathbf{J}_\parallel(t-|nz|/c)\,, \qquad z\le 0\,,
\nonumber \\
&\mathbf{E}_\mathrm{trans}(z,t)=\mathbf{E}_0(z,t)-\mu_0\frac{c}{2n}\mathbf{J}_\parallel(t-|nz|/c)\,, \qquad z\ge 0\,.
\label{eq:wave_eq-sol2}
\end{align}
The effective field that  
drives ${\bf p}_\mathrm{S}(t)$, Eq.~(\ref{ps}),
is modified as compared to the incident field by the reflected field determined by the supercurrent. 
The latter,   in turn, depends on the   displaced populations and coherences determined by ${\bf p}_\mathrm{S}(t)$. 
The effects of this  non-perturbative self-consistent 
coupling  are discussed next.

\section{Lightwave propagation effects on the non-equilibrium SC dynamics}
\label{sec:gapless}

In this section, we demonstrate that photo-excited SC nonlinearities, together with lightwave propagation inside a  SC thin film  system, can 
lead to coherent photogeneration 
of a nonlinear supercurrent with a dc component
and  a gapless moving condensate  non-equilibrium state with tunable superfluid density. 
 For our numerical calculations, we use the square lattice nearest-neighbor tight-binding dispersion $\xi(\mathbf{k})=-2\,J[\mathrm{cos}(k_x\,a)+\mathrm{cos}(k_y\,a)]+\mu$, with hopping parameter $J$, lattice constant $a$, and band-offset $\mu$. We only consider the half-filling limit  $(\mu=0)$ where particle-hole symmetry is realized. As initial state, we take  the BCS  ground state with SC gap $2\Delta=5.1$\,~meV. To compute the dynamics of the gauge-invariant density matrix (\ref{eq:rho_FT})
without perturbative expansions,  
  we self-consistently solve the SC Bloch equations (\ref{eq:eoms}) and the equations for the SC gap and the Fock energy (\ref{eq:SC-homo}) together with Eq.~(\ref{eq:wave_eq-sol}), using a fourth-order Runge--Kutta method. We excite the system with external THz  electric field $\mathbf{E}_\mathrm{THz}(t)=E_0\,\mathbf{e}_x\,\mathrm{sin}(\omega_0\,t)\,\mathrm{exp}[-t^2/(2\,\sigma_t^2)]$, where $E_0$ is the field amplitude that defines its strength, $\omega_0$ corresponds to the central frequency, and $\sigma_t$ determines the duration of the applied $\mathbf{E}$-field. The external  pump $\mathbf{E}$-field satisfies the condition $\int_{-\infty}^\infty\mathrm{d}t\,\mathbf{E}_\mathrm{THz}(t)=0$, i.e. $\mathbf{E}_\mathrm{THz}(t)$ does not contain any zero-frequency dc component. This is a  condition which every physical source of electromagnetic waves must   satisfy according to Maxwell's equations.

\begin{figure}[!tbp]
	\centering
		\includegraphics[scale=0.36]{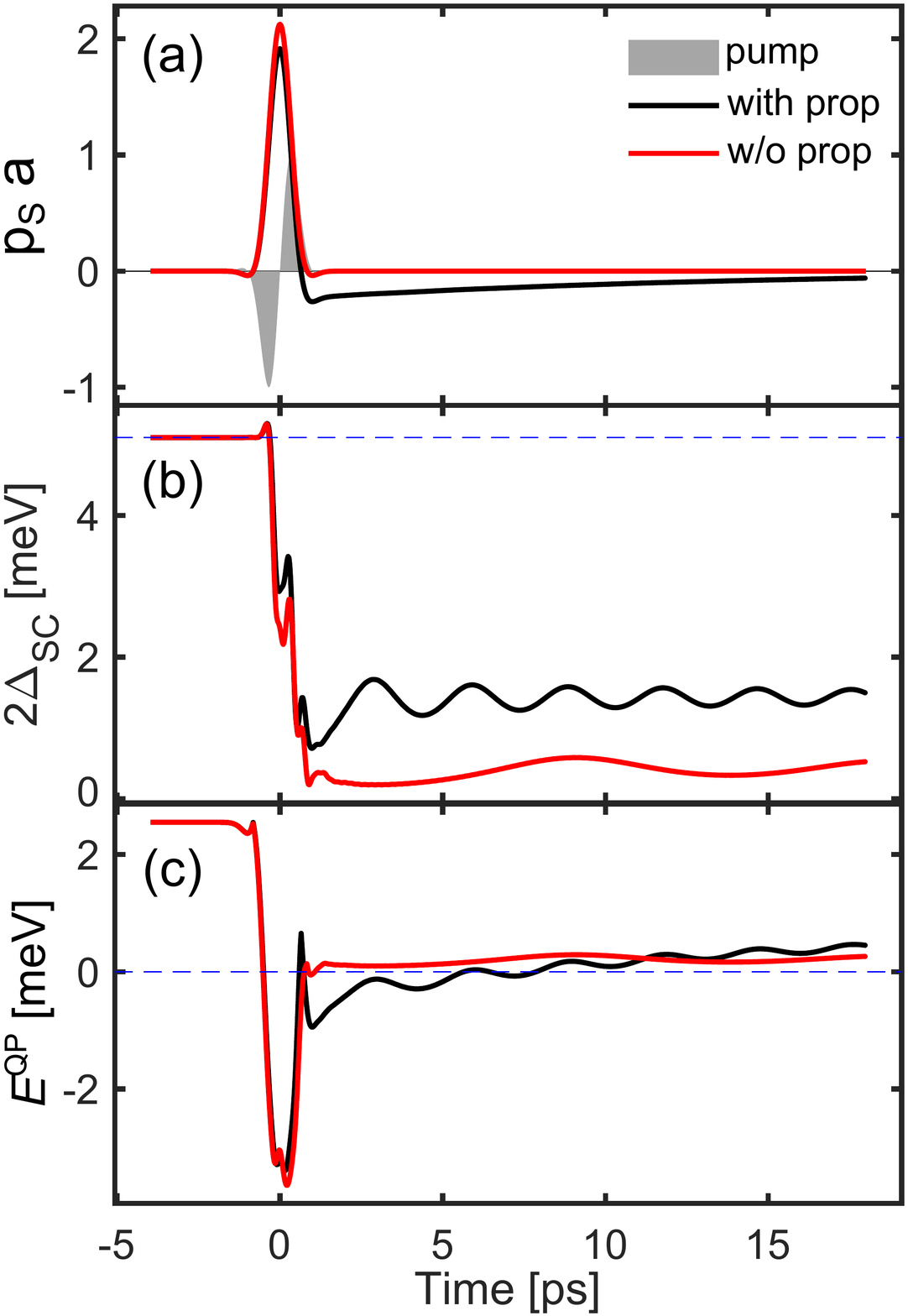}
		\caption{Time evolution of BCS state into a gapless non-equilibrium SC state after strong THz quantum quench. (a) Dynamics of THz-light-induced superfluid momentum $p_\mathrm{S}(t)$. We compare the  calculations with electromagnetic propagation effects (black line, effective driving field differs from external  field) and without (red line, driving field coincides with external field), together with a representative single-cycle 0.5~THz pump electric field  (shaded area). 
The effective driving field, determined self-consistently by Maxwell's equations and the nonlinear supercurrent, accelerates the condensate, with  a center-of-mass momentum that decays slowly  in time.
		(b) The corresponding dynamics of the SC order parameter; horizontal dashed line indicates the equilibrium value of the SC order parameter. 
The condensate density remains finite after  the pulse when  electromagnetic propagation effects 	are included.
		(c) The corresponding dynamics of the minimum QP excitation energy.
A gapless QP anisotropic spectrum with finite order parameter is obtained after the pulse when electromagnetic propagation effects are included. 	
		}
		\label{fig2} 
\end{figure}

Figure~\ref{fig2}(a) shows the dynamics of the superfluid momentum $\mathbf{p}_\mathrm{S}(t)$ driven by a short 0.5~THz pulse (shaded area). 
We compare our  calculations without (red line) and with electromagnetic propagation effects (black line).
The latter leads to an effective electric  field whose temporal profile differs from the external THz  pulse.  The SC  time evolution  is driven  by this effective  field, which depends self-consistently on the nonlinear photocurrent.
Such excitation of the SC system induces a superfluid momentum during the pulse, which persists after the pulse only when the electromagnetic propagation effects are included. 
The condensate  center-of-mass momentum  decays in time  
due to radiative damping, which is a consequence of the self-consistent coupling between the nonlinear photocurrent and the lightwave $\mathbf{E}$-field. The calculated momentum relaxation rate $\Gamma$, derived in Appendix~\ref{app:damping}, depends on the details of the bandstructure and is stronger for superconductors with a large density of states at the Fermi surface~\cite{vaswani2019discovery}.  

Due to the condensate motion, ${\bf p}_\mathrm{S} \ne 0$, the SC order parameter 
2$\Delta$ 
no longer coincides with the QP excitation gap. 
The dynamics of $\Delta(t)$ and the QP excitation energy are plotted in Figs.~\ref{fig2}(b) and (c). Here the QP excitation energy  is defined as the minimum of the QP energy $E^\mathrm{QP}_\mathbf{k}$ among all $\mathbf{k}$ around the Fermi surface, with $E^\mathrm{QP}_\mathbf{k}$  given by Eq.~(\ref{eq:Eqp}). The results of our calculations with or without electromagnetic  propagation effects both show a quench of the SC order parameter followed by Higgs oscillations. At the same time, the QP excitation spectrum can transiently become  gapless during the pulse for sufficiently high fields, i.e. $E^\mathrm{QP}\le 0$ for some $\mathbf{k}$. However,  for the calculation where electromagnetic propagation effects are included, we obtain a non-equilibrium state where 
 the QP excitation spectrum is also gapless after the pulse for sufficiently strong 
 fields. 
Despite this gapless excitation spectrum,   
  the SC order parameter remains finite,
which corresponds to a gapless condensate non-equilibrium state
consistent with recent experimental observations~\cite{yang2019lightwave}. 
 The above controllable gapless 
non-equilibrium quantum state   
 arises  from  THz dynamical symmetry breaking 
in a  moving condensate,  which is absent for the standard Anderson pseudo-spin model. With lightwave acceleration during cycles 
of carrier wave oscillations,
we  can thus non-adiabatically drive a gapless SC quantum phase with finite condensate coherence in a SC thin film. For this quantum state,   $\Delta >0$ while $E^\mathrm{QP}_\mathbf{k} \le 0$ for several $\mathbf{k}$-points.  For the same pump electric field, a quenched SC state with $\Delta>0$ and $E^\mathrm{QP}_\mathbf{k}>0$ is obtained if the  electromagnetic propagation effects are neglected.

\begin{figure}[!tbp]
	\centering
		\includegraphics[scale=0.36]{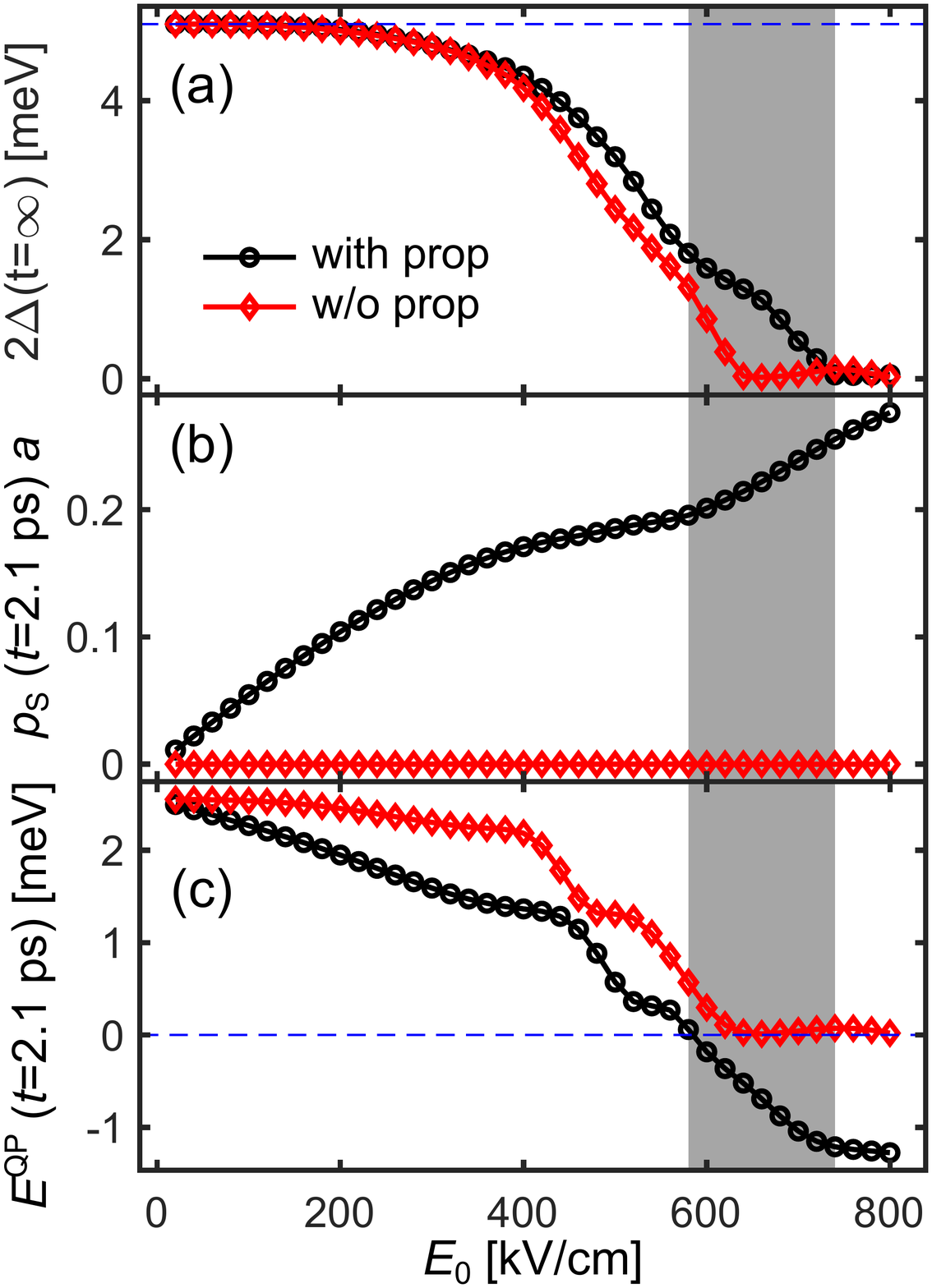}
		\caption{Control of THz-driven non-equilibrium SC state by pump electric field  strength. (a) Pump field amplitude  dependence of the steady state SC order parameter reached after the pulse.  We compare  the  calculations with electromagnetic propagation effects (black line, effective driving field differs from  external pulse) and without (red line, driving field coincides with  external THz laser pulse).  Shaded area indicates phase (II) with finite condensate density but gapless excitation spectrum. (b), (c) The corresponding $E_0$-field dependence of the superfluid momentum and QP excitation energy.}
		\label{fig3} 
\end{figure}

\begin{figure*}[!tbp]
	\centering
		\includegraphics[scale=0.39]{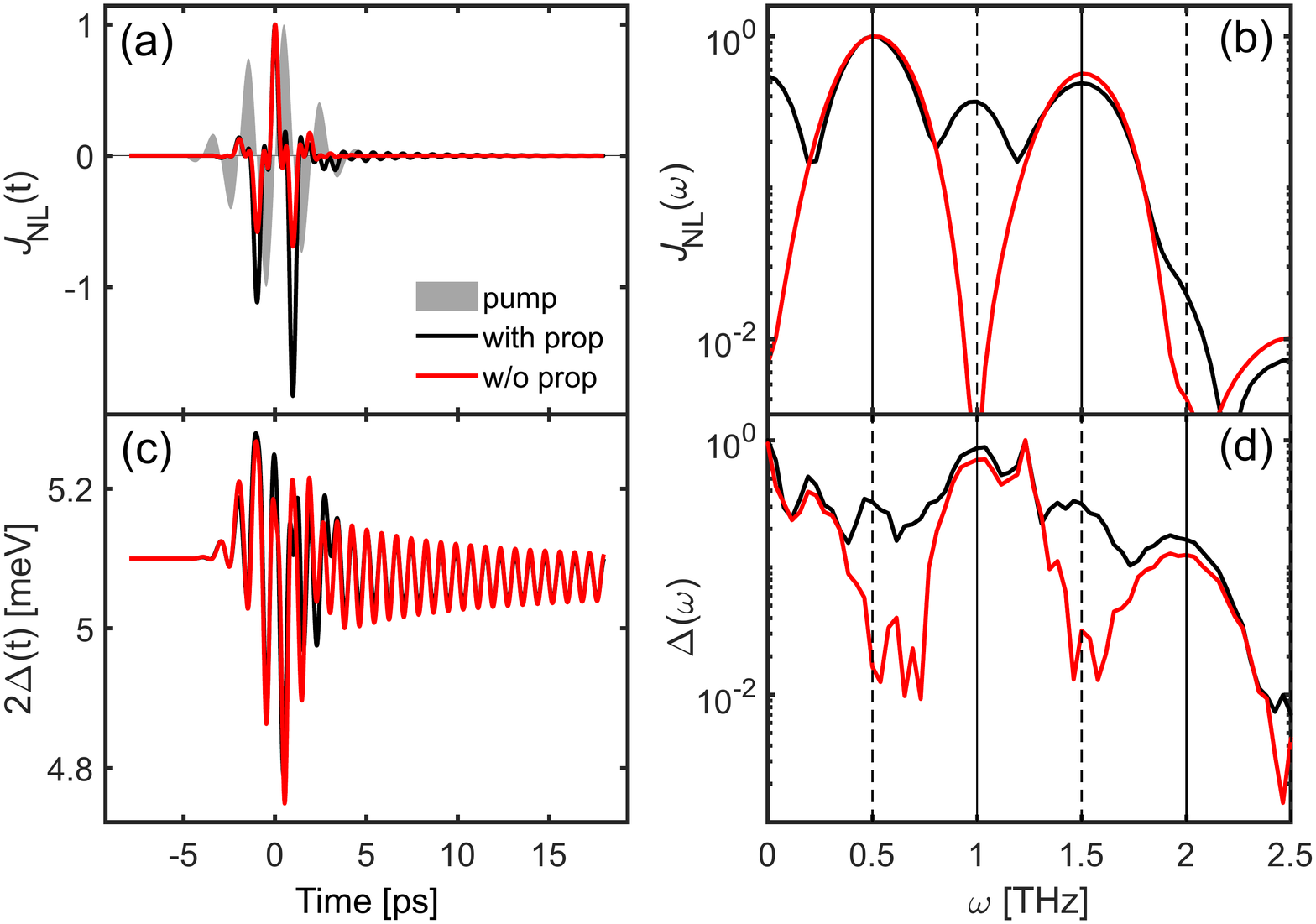}
		\caption{Detection of THz dynamical inversion symmetry breaking by HHG. (a) Dynamics of THz-light-induced nonlinear current $\mathbf{J}_\mathrm{NL}(t)$ for a calculation with (black line) and without electromagnetic propagation effects (red line), shown  together with a  representative 0.5~THz electric pump field used in the calculations (shaded area). (b) The corresponding supercurrent spectra in semi-logarithmic scale. Vertical solid (dashed) lines indicate equilibrium-symmetry allowed (forbidden) harmonics. (c), (d) The corresponding dynamics and spectra of the SC order parameter.}
		\label{fig4} 
\end{figure*}

Figure~\ref{fig3} illustrates our conclusion that THz lightwave propagation inside the SC system can selectively drive  three different non-equilibrium quantum phases (I)-(III) after the pulse. This result  opens up possibilities for coherent control, as the calculated  light-tunable transient quantum states define a systematic  initial condition for  post-quench long-time dynamics that is consistent with recent experiments~\cite{yang2018,vaswani2019discovery}. The $E_0$-field dependence of the steady state SC order parameter reached after the pulse (Fig.~\ref{fig3}(a)), the laser-induced superfluid momentum (Fig.~\ref{fig3}(b)), and the QP excitation energy slightly after the pump pulse (Fig.~\ref{fig3}(c)) are shown for a calculation with (black line) and without  (red line) propagation effects. Without lightwave propagation inside the SC system, increasing the pump field amplitude $E_0$ can only drive  a quenched SC state with $\Delta>0$ and $E^\mathrm{QP}>0$ (regime I) or a gapless QP state  with $\Delta=0$ and $E^\mathrm{QP}=0$ (regime III). In this case,  $\mathbf{p}_\mathrm{S}\neq 0$ only during the 
electromagnetic pulse. 
The steady states after the pulse  are then  similar to the quantum states obtained after  ``sudden quench" of the order parameter~\cite{Yuzbashyan:2006,Forster2017}. 
Compared to that, photo-excited nonlinearities together with lightwave propagation inside the superconductor  lead to a finite center-of-mass momentum of the  accelerated condensate after the pulse, which can drive a gapless SC state with finite coherence (order parameter) across a wide range of $E_0$ (shaded area, regime II). This result  is consistent with the experimental observations in Ref.~\cite{yang2019lightwave} and cannot be obtained by using 
the standard Anderson pseudo-spin Hamiltonian without including the lightwave quantum transport effects. 
Lightwave sub-cycle acceleration results in an oscillating condensate 
center-of-mass momentum $\mathbf{p}_\mathrm{S}$ that  remains finite after 
the pulse due to THz dynamical symmetry breaking
in a  thin film geometry. 
This   lightwave  acceleration modulates the SC excitation spectrum, which 
can transiently  close and reopen during cycles of  carrier wave oscillations. 
In addition to coherently controlling gapless non-equilibrium SC, 
the above THz dynamical symmetry breaking and coherent nonlinear supercurrent photogeneration manifest themselves 
in HHG at equilibrium-symmetry-forbidden frequencies, discussed next.

\section{Experimental detection of lightwave dynamical symmetry breaking:  coherent control of  high-harmonics generation
\label{sec:HHG}}

As shown in the previous section,  photo-excited SC nonlinearities together with lightwave propagation effects can lead to coherent photogeneration of a nonlinear supercurrent with a $\omega \approx 0$ component. 
In addition to driving gapless SC states after the pulse, such 
THz dynamical inversion symmetry breaking  allows us to coherently 
control  HHG in the nonlinear response via  the 
momentum and excitations  of the  moving condensate  state
  during cycles 
of carrier wave oscillation. 
  For  a driving field 
with $\int_{-\infty}^\infty\mathrm{d}t\,\mathbf{E}_\mathrm{THz}(t)=0$,  the condensate center-of-mass momentum $\mathbf{p}_\mathrm{S}(t)$ oscillates symmetrically in time  with the pump laser's frequency $\omega_\mathrm{0}$.
 $\mathcal{O}(\mathbf{p}_\mathrm{S}^2)$ terms in the equations of motion (\ref{eq:eoms}) then drive a temporal evolution of the density matrix $\tilde{\rho}(\mathbf{k})$ with symmetry-allowed frequency oscillations at $2\omega_\mathrm{0}$. The  nonlinear contributions to the equations of motion also produce  higher {\em even} harmonics $4\omega_\mathrm{0},\,6\omega_\mathrm{0},\dots$ in the density matrix time evolution. As a result, the current $\mathbf{J}$ (order parameter $\Delta$) shows odd (even) harmonics of the pump laser pulse's  frequency. The above situation changes, however,  when electromagnetic  propagation effects are included to obtain the real driving  field.  As 
discussed in the previous sections, the latter can lead to  inversion symmetry breaking
 of the electron and hole distributions in momentum space that persists after the pulse.  The spectrum of $\mathbf{p}_\mathrm{S}(\omega)$ now shows a  small zero-frequency light-induced nonlinear component,  in addition to the peak at $\omega_0$.  Such dc contribution results from  the effective electric field
that accelerates ${\bf p}_\mathrm{S}(t)$,  
   which  is modified from the external field  by the  oscillating nonlinear supercurrent.  As a result of THz dynamical symmetry breaking, 
 the Fourier transformation of $\mathbf{J}$ ($\Delta$) will exhibit equilibrium-symmetry forbidden even (odd) harmonics, in addition to the well-known odd (even) harmonics. While the Anderson pseudo-spin model predicts third harmonic generation  studied in the past,   THz dynamical symmetry breaking 
during cycles of lightwave oscillations leads to forbidden HHG modes recently observed experimentally~\cite{vaswani2019discovery}.

To test the above perspective, we plot in Fig.~\ref{fig4}(a) the dynamics of THz-light-induced nonlinear supercurrent $\mathbf{J}_\mathrm{NL}(t)$ for a calculation with (black line) and without (red line) electromagnetic propagation effects. Here,  the SC system is excited with a 8~ps THz pulse (shaded area). Since a linear photocurrent only produces a $\omega_\mathrm{0}$-frequency contribution, we focus on the nonlinear photocurrent.
 The latter   exhibits third harmonic oscillations, i.~e. oscillates with $3\,\omega_0$,  higher harmonics, and  a small dc  component $J_\mathrm{dc}$ that decays slowly with a rate $\Gamma$ calculated in appendix~\ref{app:damping}. Also, $\mathbf{J}_\mathrm{NL}$ shows pronounced amplitude Higgs oscillations due to the photoinduced $J_\mathrm{dc}$ that breaks the symmetry.  A Fourier transformation of the nonlinear current temporal profile allows us to disentangle the different nonlinear optical processes contributing to $\mathbf{J}_\mathrm{NL}(t)$. The calculated emission spectrum, $I(\omega)=|\mathbf{J}_\mathrm{NL}(\omega)|^2$, at energy $\hbar\omega$ is presented in Fig.~\ref{fig4}(b) in semi-logarithmic scale. While the spectrum resulting from the calculation without propagation effects shows only odd harmonics (solid vertical lines), the result of the full calculation yields odd as well as even harmonics (dashed vertical lines). To confirm that the broken inversion symmetry of the non-equilibrium state can be detected by HHG, Figs.~\ref{fig4}(c) and (d) present the corresponding dynamics and spectra of the SC order parameter. Without electromagnetic propagation effects, $\Delta(\omega)$  only shows even harmonics, while lightwave propagation inside the SC system leads also to the generation of equilibrium-symmetry-forbidden odd harmonics. The latter demonstrates that THz dynamical inversion symmetry breaking induced by nonlinear supercurrent coherent photogeneration is directly detectable by HHG. Such nonlinear response presents a direct experimental verification of the theoretically predicted effect,  confirmed in Ref.~\cite{vaswani2019discovery}, and can be coherently controlled by tuning the applied  few-cycle field.

\begin{figure}[!tbp]
	\centering
		\includegraphics[scale=0.40]{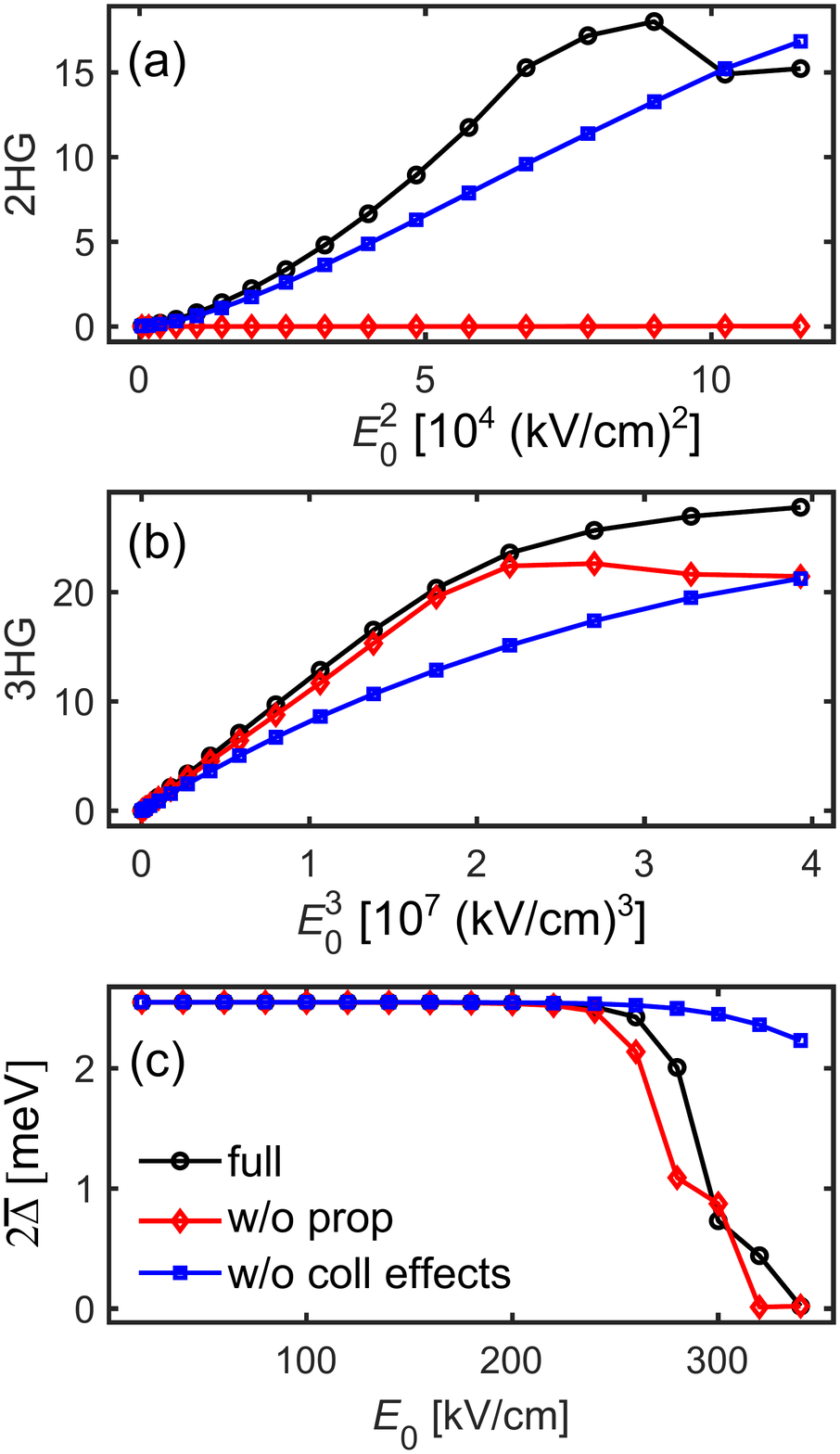}
		\caption{Electric field strength $E_0$ dependence of HHG. (a) Second harmonic emission as a function of electric field strength; the full calculation (black line) is compared with a calculation where electromagnetic propagation effects (red line) and light-induced changes in the collective effects (blue line) are switched off. (b), (c) The corresponding $E_0$-dependence of third harmonic generation and the average value of the SC order parameter in the steady state after the THz pulse, $\bar{\Delta}$.}
		\label{fig5} 
\end{figure}

We next investigate the electric field strength dependence of HHG and identify the nonlinearities contributing to each different  HHG peak by applying a switch-off analysis. Figures~\ref{fig5}(a) and (b) show the electric field strength dependence of second (forbidden) harmonic and third harmonic generation, respectively, while the average value of the SC order parameter in the steady state after the THz pulse,  $\bar{\Delta}$, is plotted in Fig.~\ref{fig5}(c). The full calculation (black line) is compared with a calculation where electromagnetic propagation effects (red line) and light-induced changes in collective effects (blue line) are switched off. The role of light-induced collective effects and corresponding interaction-induced nonlinearities is revealed by replacing $\Delta(t)$ by the equilibrium SC gap $\Delta_0$ on the right-hand side of the equation of motions~\cite{Anderson}. 
In this case,  the fluctuations, $\delta\Delta(t)=\Delta(t)-\Delta_0$, of the SC order parameter are neglected, so the response is fully determined by charge-density fluctuations. For the full calculation, the emitted intensity of second (third) harmonics grows linearly as a function of $E_0^2$ ($E_0^3$) at low pump fields, before saturating at elevated $E_0$, where the SC order parameter becomes quenched (Fig.~\ref{fig5}(c)). While switching off propagation effects only slightly reduces the  third harmonic emission, second harmonic emission is zero in this case, due to persisting inversion symmetry.
 Compared to charge fluctuations, collective effects 
 affect the nonlinear emission  in the non-perturbative regime, where one observes deviations from the linear behavior expected from susceptibility expansions.  
In particular, collective effects in the order parameter time-dependence,
coming from the light-induced $\delta\Delta(t)$, 
 enhance  both the second and third harmonic emission in the nonlinear regime. However, 
 charge fluctuations dominate 
 in the linear regime described by susceptibility 
expansions, as in earlier studies ~\cite{Cea2016}. 
For such small fields,   the quench of the SC order parameter from  equilibrium is small 
(Fig. 3(c)). 
 The effect of Fermi sea pockets on the above result will be discussed elsewhere. 

\begin{figure}[!tbp]
	\centering
		\includegraphics[scale=0.40]{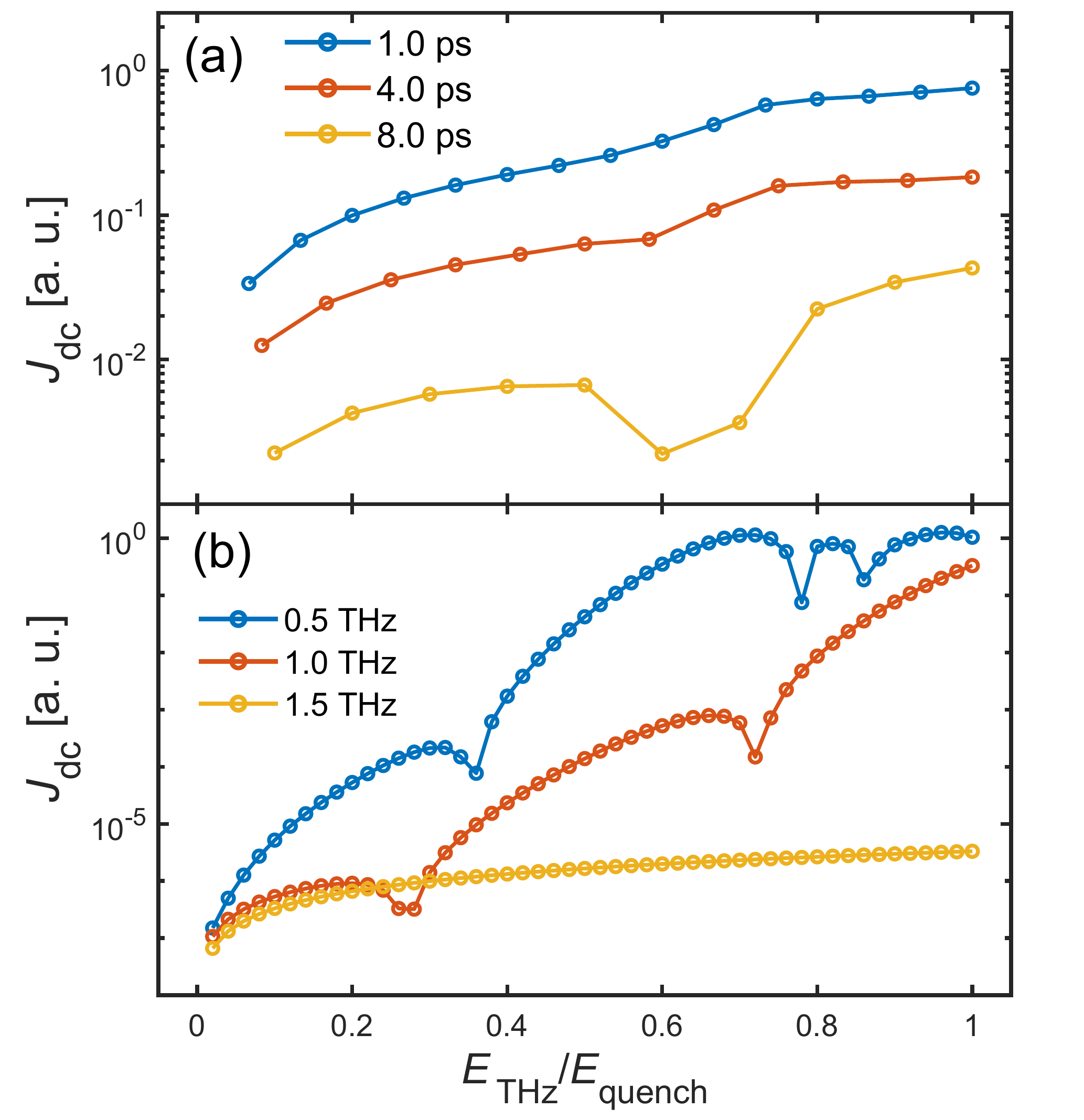}
		\caption{Coherent control of nonlinear dc supercurrent photogeneration. (a) Electric field strength dependence of the THz light-induced dc supercurrent for three different pulse durations with fixed frequency, for fields up to complete quench of the SC order parameter.  $E_\mathrm{quench}$ denotes the peak electric field needed to completely quench $\Delta$. (b) Electric field strength dependence of $J_\mathrm{dc}$ for three different pump frequencies with fixed duration.}
		\label{fig6} 
\end{figure}

Figure~\ref{fig6} demonstrates the nonlinear origin of the
symmetry-breaking  $J_\mathrm{dc}$. The latter  can be controlled by tuning the multi-cycle THz field temporal profile, i.e., the number of cycles of oscillation  and  frequency $\omega_0$. Figure~\ref{fig6}(a) shows the pump fluence-dependence of the THz-lightwave-induced dc supercurrent for three different pulse durations with fixed $\omega_0$. 
We consider field strengths up to  complete quench of the SC order parameter  ($E_\mathrm{quench}$).   The photoinduced $J_\mathrm{dc}$, which characterizes  the THz dynamical symmetry breaking,
 increases with decreasing pulse duration, i.e., with  fewer  cycles
 of oscillations. In this case,  $\int_{-\infty}^\infty\mathrm{d}t\,J(t)$ is larger such that the forward and backward propagating electromagnetic fields (\ref{eq:wave_eq-sol}) inside the SC film become more asymmetric which leads to a larger $J_\mathrm{dc}$. The  pump-frequency dependence of the  supercurrent for fixed  pulse duration 
  is shown in Fig.~\ref{fig6}(b), which  demonstrates that $J_\mathrm{dc}$  is  stronger for the  lower-frequency pulses, which again corresponds to  fewer cycles of oscillation.

\begin{figure}[!tbp]
	\centering
		\includegraphics[scale=0.25]{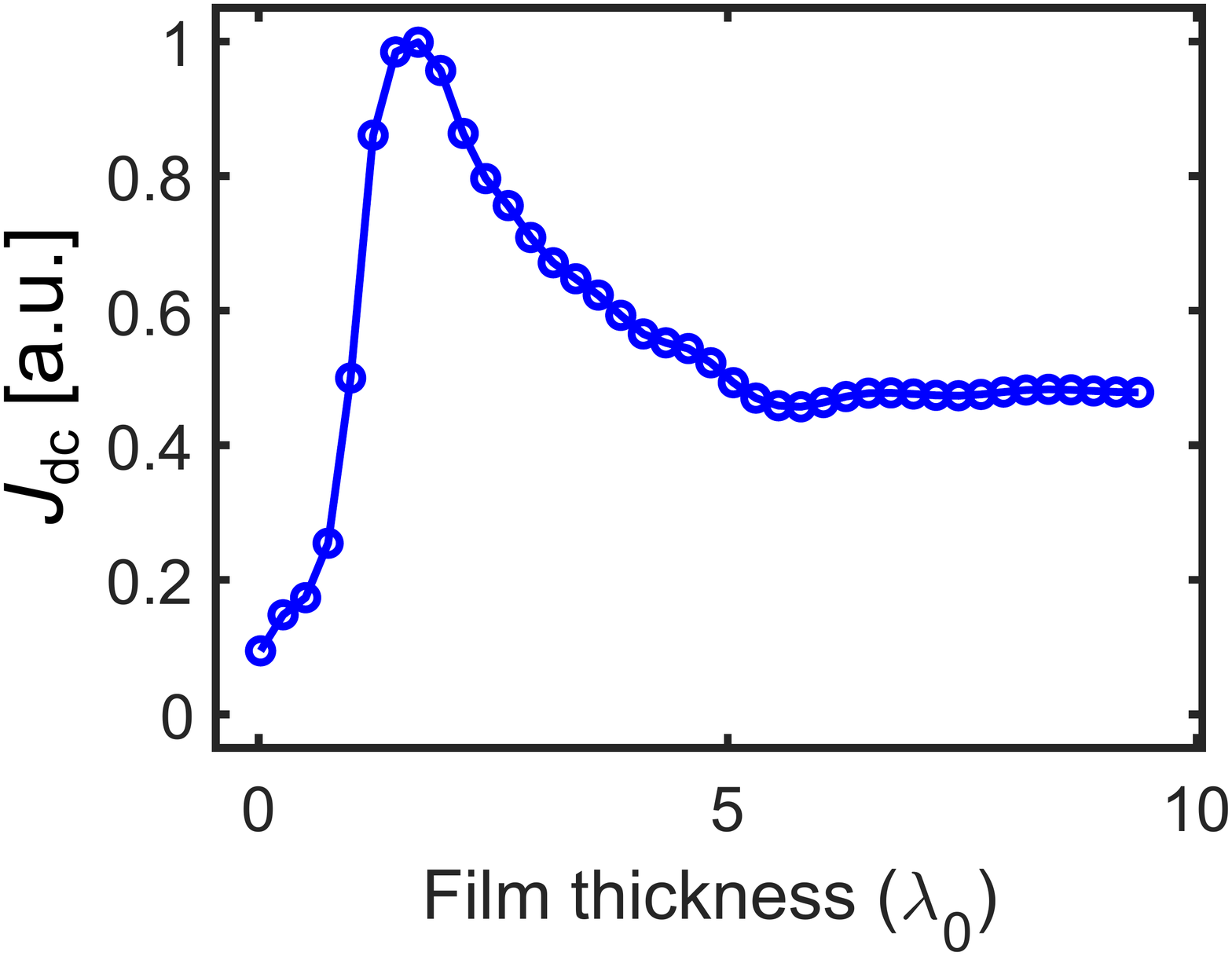}
		\caption{Effect of SC film thickness on dc supercurrent photogeneration for 20~ps-long pump pulse with  field strength of 10.0~kV/cm; $\lambda_0$ denotes the wavelength of the pump.}
		\label{fig7} 
\end{figure}

The photogeneration of  dc nonlinear supercurrent component and the resulting second harmonic symmetry-forbidden light emission also depend on the    bandstructure, especially on the density of states,
and can also be controlled by varying the thickness of the SC film~\cite{vaswani2019discovery}. The film thickness dependence of $J_{dc}$ 
is illustrated in Fig.~\ref{fig7}.  The latter dependence is dominated by the  interference inside the SC film of the light-induced nonlinear current, the incident $E$-field, and the reflected  $E$-field,   analogous  to four-wave mixing. For small film thicknesses, $J_\mathrm{dc}$ grows nonlinearly up to roughly 2$\lambda_0$, where $\lambda_0$ is the wavelength of the pump. In this regime, the region within the SC film where all three waves interfere during the nonlinear dynamics increases with thickness, which results in the increase of the dc supercurrent. With$  $ increasing film thickness, the time delay between current and reflected THz lightwave field grows. As a result, the interference between both fields is reduced in some regions within the SC film,  due to radiative damping of the current. At the same time, the photoinduced dc supercurrent quenches the SC order parameter. Both effects lead first to a decrease of the dc photocurrent, before  $J_\mathrm{dc}$ saturates at larger film thicknesses.

\section{Nonlinear collective mode phase-coherent spectroscopy}
\label{sec:Higgs}

\begin{figure}[!tbp]
 \centering
		\includegraphics[scale=0.41]{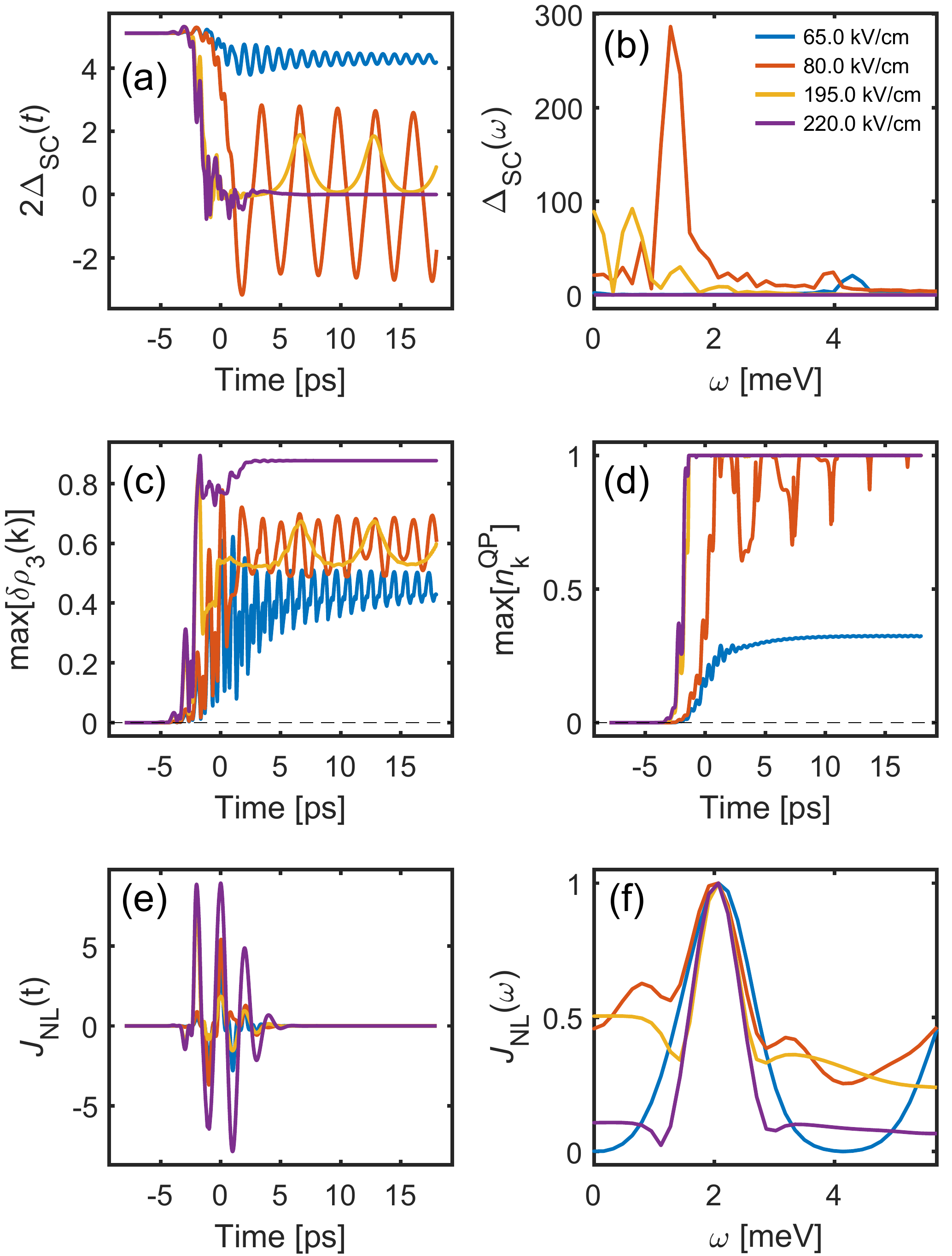}
		\caption{Selective driving of different collective modes of the SC order parameter. (a), (b) Dynamics and spectra of the SC order parameter for various electric field strengths. (c), (d), (e) The corresponding dynamics of the maximum of $\delta\tilde\rho_3(\mathbf{k})$ among all wavevectors $\mathbf{k}$, the maximum of the QP distributions among all $\mathbf{k}$, and the nonlinear current. (f) The corresponding spectra of the nonlinear current.}
		\label{fig8} 
\end{figure}

In this section, we focus on  the collective mode dynamics of  the THz-light-driven SC
state. We  explore, in particular,  ways in which we can  control the various 
dynamical phases that can evolve in time 
from the SC ground state   by tuning  the interplay of THz quantum quench, lightwave condensate acceleration, and THz dynamical symmetry breaking. 
Figure~\ref{fig8}(a) shows the changes in the time evolution of the SC order parameter for various pump fields. To characterize the oscillations, 
we plot the corresponding spectra obtained by Fourier transformation  
of the order parameter time dependence   in Fig.~\ref{fig8}(b). 
Here we show results when the system is driven by  a  few-cycle 0.5~THz pulse
which, as discussed in the previous section,  maximizes the $\omega=0$ component of the nonlinear  supercurrent. 
Figure~\ref{fig8}(a) demonstrates  
 four different amplitude collective  modes selectively excited by such field: (I) In the low excitation regime, $\Delta(t)$ shows damped oscillations (blue line) with frequency $2\Delta_\infty > 0$.
 This Higgs amplitude mode
\cite{Axt2007,Yuzbashyan:2006,Forster2017} 
decays as $t^{-1/2}$ (for one-band superconductors)  to the steady state order parameter value $\Delta_\infty$, as a 
  result of Landau damping due to energy transfer of the collective mode to QPs.   $2\Delta_\infty=4.2$~meV  coincides with  the position of the peak in the $\Delta(\omega)$ spectrum (blue line in Fig.~\ref{fig8}(b)). (II) By 
increasing the field strength until we quench 
the order   parameter,  $\Delta_{\infty} \approx 0$,
we obtain a number of highly nonlinear gapless dynamical phases, whose behavior changes  by varying the field strength (Rabi energy) 
as well as the cycles of oscillation. In regime II, 
 the order parameter displays  strong  persistent oscillations around the steady state value $\Delta_{\infty}$=0,  with multiple frequencies (red line, Fig.~\ref{fig8}(b)). 
 This undamped collective mode is due to a synchronization of QP Rabi oscillations 
 excited by the  pulse,
  in a 
non-equilibrium quantum state with $\Delta_{\infty} =0$ but 
with a  time-dependent coherence
\cite{soliton,Balseiro}. 
 (III) Further increase of the pump field amplitude in this highly nonlinear regime with finite condensate momentum leads to excitation of anharmonic damped oscillations
of a finite order parameter
  (yellow line, Fig.~\ref{fig8}(a)). The corresponding spectrum (yellow line, Fig.~\ref{fig8}(b)) shows a main peak at $0.7$~meV, while high  harmonics arise due to the anharmonicity of the order parameter oscillation. This new mode  is a consequence of the  anisotropic electron and hole distributions
driven by the lightwave condensate  acceleration,
which are displaced in ${\bf k}$ space by ${\bf p}_\mathrm{S}(t)$. 
Such  dynamical phase is not accessible by  the isotropic  order parameter sudden quench or by using the standard Anderson pseudo-spin precession models.  In our theory,  the 
electronic distribution  is angle-dependent due to  THz dynamical symmetry breaking determined by the direction of ${\bf p}_\mathrm{S}(t)$, 
which can be controlled by the lightwave  polarization.
 (IV) In the extreme nonlinear regime, the dynamics becomes over-damped and the order parameter decays exponentially  to  $\Delta_\infty=0$ (purple line, Fig.~\ref{fig8}(a))~\cite{dzero,sychron}. 
 We conclude that THz dynamical symmetry breaking and ${\bf k}$ space 
 anisotropy controlled via lightwave condensate acceleration allows us to  selectively drive  dynamical phases  (II) and (III), in addition to 
the familiar phases (I) and (IV).
While Phase (II) is also accessible
by periodic modulation of Hamiltonian parameters~\cite{Balseiro},  
here we obtain multiple frequencies due to the anisotropic 
distributions induced by ${\bf p}_\mathrm{S}(t)$. 
Furthermore, 
the  time evolution  depends on the 
synchronization between the pump and Rabi oscillations. 

\begin{figure*}[!tbp]
 \centering
		\includegraphics[scale=0.4]{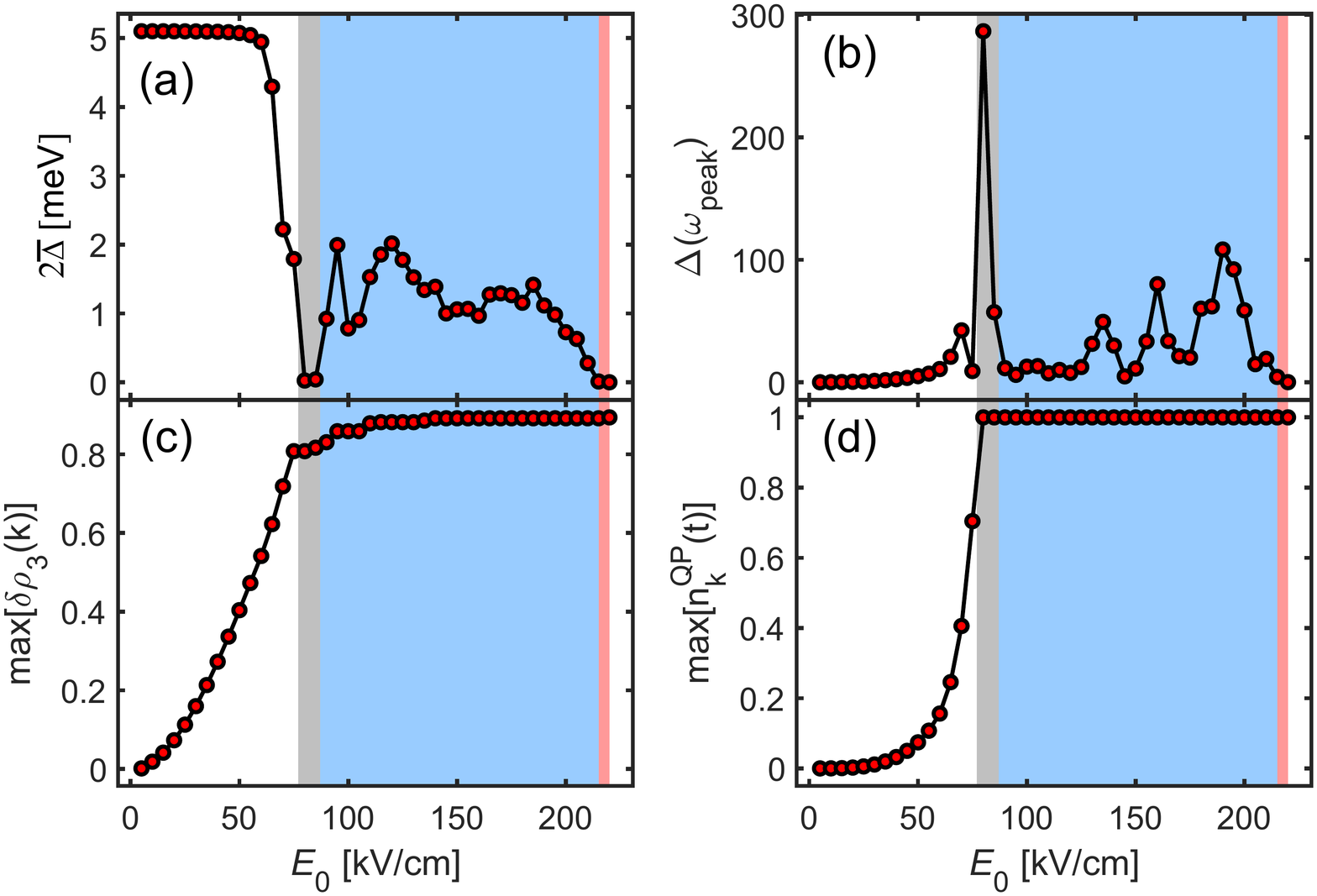}
		\caption{Control of collective amplitude modes by the pump field. Electric field strength dependence of (a) the average value of the order parameter in the non-equilibrium state, $\bar{\Delta}$, (b) the oscillation amplitude of the main peak in $\Delta(\omega)$ spectra, $\Delta(\omega_\textbf{peak})$, (c) population inversion $\textrm{max}[\delta\tilde\rho_3(\mathbf{k})]$, and (d) quasi-particle distribution $\textrm{max}[n^\textrm{QP}_\textbf{k}]$. Non-equilibrium phase (II) ((III)) is indicated by a gray (blue) shaded area, while phase (IV) is denoted by a red shaded area.}
		\label{fig9} 
\end{figure*}

To investigate the systematic lightwave driving  of the  different non-equilibrium quantum phases, we study the dynamical change of the population inversion $\delta\tilde\rho_3(\mathbf{k})$, defined in Eq.~(\ref{eq:rho_ps}),
in more detail. Figure~\ref{fig8}(c) shows the dynamics of the maximum of $\delta\tilde\rho_3(\mathbf{k})$ among all wavevectors $\mathbf{k}$ for the four different dynamical phases  in Fig.~\ref{fig8}(a). We observe strong population inversion oscillations during the pulse, i.e., Rabi--Higgs oscillations~\cite{Balseiro}. This population inversion occurs during cycles of lightwave oscillations and defines the initial condition during the pulse for driving collective mode dynamics after the pulse. While for mode (I) $\delta\tilde\rho_3(\mathbf{k})<0.5$ is small after the pulse,
in which case we recover previous non-equilibrium states also obtained within the Anderson pseudo-spin model, 
phases (II)--(IV) emerge in the extreme nonlinear excitation regime, where the pseudo-spin populations are inverted as compared to the BCS ground state, i.e. $\delta\tilde\rho_3(\mathbf{k})>0.5$ for some $\mathbf{k}$. 
To study the  emergence of these 
light-induced dynamical phases 
by controlling the population inversion in more detail, 
 Fig.~\ref{fig8}(d) shows the dynamics of the maximum of the QP distributions among all $\mathbf{k}$, $\textrm{max}[n^\textrm{QP}_\textbf{k}]$, for the four  different phases. For (II)--(IV), THz excitation has created large QP populations with QP distributions close to one, i.e. $\rho_{1,1}^\mathrm{qp}(\mathbf{k})=\rho_{2,2}^\mathrm{qp}(\mathbf{k})=1$ for some $\mathbf{k}$. These inverted populations remain fully occupied after the pulse for modes (III)--(IV). In particular, 
as a result
of THz dynamical symmetry breaking by ${\bf p}_\mathrm{S}(t)$,  
 $\mathbf{k}$-space is separated into 
two coexisting regions: 
a SC region 
and a blocking region.
The latter consists of ${\bf k}$ points 
 where Cooper pairs are broken and QP states are fully occupied. 
 The blocking region of ${\bf k}$-space leads to a strong 
suppression of the anomalous expectation values and the SC order parameter
 determined by ${\bf p}_\mathrm{S}(t)$, 
 such that excitation of collective modes (II)--(IV) 
becomes possible. The latter 
is achieved 
via  the anisotropy in ${\bf k}$-space introduced  by THz dynamical symmetry breaking for a moving condensate, 
which results in multiple frequencies for (II) and (III). 
 The dynamical quantum phases driven by the pronounced Rabi--Higgs oscillations modify the light emission spectrum, 
which makes them experimentally observable.  
  The dynamics and spectra of the nonlinear current $J_\mathrm{NL}$ for the four different amplitude modes are shown in Figs.~\ref{fig8}(e) and (f). The collective modes (II) and (III) lead to sideband generation around the fundamental harmonic at $\omega_0=0.5$~THz. The energy of these sidebands matches the fundamental frequency of the dynamical mode observable in the $\Delta(\omega)$-spectra in Fig.~\ref{fig8}(b), so the predicted quantum phases are detectable by looking at the emission spectrum and are controlled  by  ${\bf p}_\mathrm{S}(t)$. 

The pump field dependence of the above  driven non-equilibrium phases is analyzed in Fig.~\ref{fig9}. There we show as a function of
the  pump field (a) the average value of the order parameter in the non-equilibrium state, $\bar{\Delta}$, and (b) the oscillation amplitude of the main peak in the $\Delta(\omega)$ spectra, $\Delta(\omega_\textbf{peak})$.  For low pump fields $E_0<75$~kV/cm, 
i.e. prior to strong SC quench, 
the SC order parameter shows damped oscillations with frequency $2\Delta_\infty=2\bar{\Delta}$ (phase (I)). At the same time, the oscillation amplitude monotonically increases analogous to the interaction quench result~\cite{Yuzbashyan:2006}. 
With increasing $E_0$, the system enters dynamical phase (II) (gray shaded area),  where $\bar{\Delta}=0$ (quenched SC order) but $\Delta(\omega_\mathrm{peak})>0$ (quantum fluctuations). Here, the order parameter shows the strongest oscillation amplitude (quantum fluctuations), i.e 
the collective mode is amplified
(see gray area in Fig.~\ref{fig9}(b)).
 A further increase of $E_0$ leads to anharmonic damped oscillations (phase (III), (blue shaded area)) across a wide range of pump fields. For $E_0$ exceeding $215$~kV/cm, the system evolves towards phase (IV) (red shaded area) after THz-driven quench where $\bar{\Delta}$=$\Delta(\omega_\textbf{peak})$=0.  The corresponding field-dependence of the population inversion and QP occupations after the pulse, plotted in Figs.~\ref{fig9}(c) and (d), demonstrate that the collective modes (II)--(IV) only emerge in the extreme nonlinear regime, where strong Rabi flopping $\delta\tilde\rho_3(\mathbf{k}) > 0.5$ and large QP densities $\rho_{1,1}^\mathrm{qp}(\mathbf{k})=\rho_{2,2}^\mathrm{qp}(\mathbf{k})=1$ are present.

\begin{figure}[!tbp]
 \centering
		\includegraphics[scale=0.4]{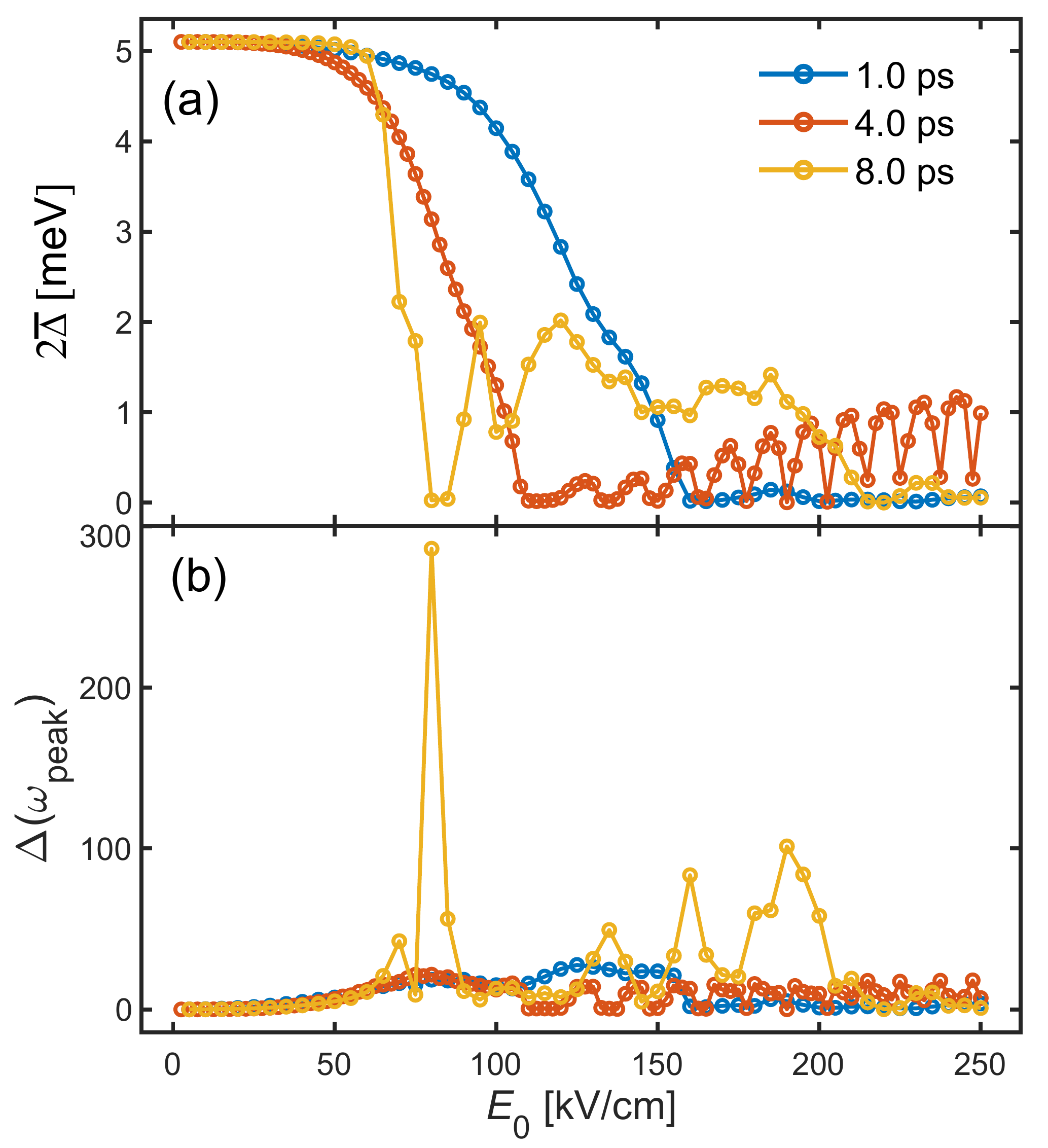}
		\caption{THz temporal profile dependence of driven collective mode phases. Electric field strength dependence of (a) the average value of the order parameter in the non-equilibrium state, $\bar{\Delta}$, and (b) the oscillation amplitude of the main peak in $\Delta(\omega)$ spectra, $\Delta(\omega_\textbf{peak})$, for three different pump pulse durations.}
		\label{fig10} 
\end{figure}

The collective  modes of the SC order parameter are not only controllable by the pump field strength as above,  but also by its temporal profile, 
i.e., by  the cycles of oscillation (frequency and duration) that determine the electromagnetic driving. 
This is illustrated in Figs.~\ref{fig10}(a) and (b), where the pump field dependence of $2\bar{\Delta}$ and $\Delta(\omega_\textbf{peak})$ are shown for  different number of cycles, obtained by fixing the frequency 
and varying the  excitation duration: 
 $1$~ps (blue line), $4$~ps (red line), and $8$~ps (yellow line). For short pump driving (blue line), the system cannot perform a full Rabi flop
required for phases (II) and (III).  
 In this case,  one can only access phases (I) and (IV) similar to the  sudden quench of the SC order parameter. 
 However, in contrast to sudden quench, $\Delta(\omega_\textbf{peak})$ does not monotonically increase 
 within phase (I) up to the transition to phase (IV)
(blue line in  Fig.~\ref{fig10}(b)).  
In particular, $\Delta(\omega_\textbf{peak})$ starts to saturate and then slightly decreases when the pump pulse frequency becomes resonant to  
 $2\Delta(t)$, as the latter  deviates  from its equilibrium value with time
 in the non-perturbative regime. 
  Close to the  transition to phase (IV), $\Delta(\omega_\textbf{peak})$ grows again, before dropping to zero in phase (IV). 
  The situation changes for intermediate excitation durations  
  with increasing number of cycles, in which case phase (III)  becomes accessible 
via Rabi--Higgs oscillations. 
  More specifically, we obtain several transitions between phases (III) and (IV) in the extreme nonlinear regime (red curve). 
  For pulse durations longer than one Rabi flop (8~ps, yellow line), one can also excite phase (II), by adjusting the pump electric field (yellow curve). In this case, we obtain a strong amplification of the Higgs mode strength (yellow curve in Fig.~\ref{fig10}(b))
as we quench the order parameter and transition to dynamical phase (II).

\begin{figure*}[!tbp]
 \centering
		\includegraphics[scale=0.45]{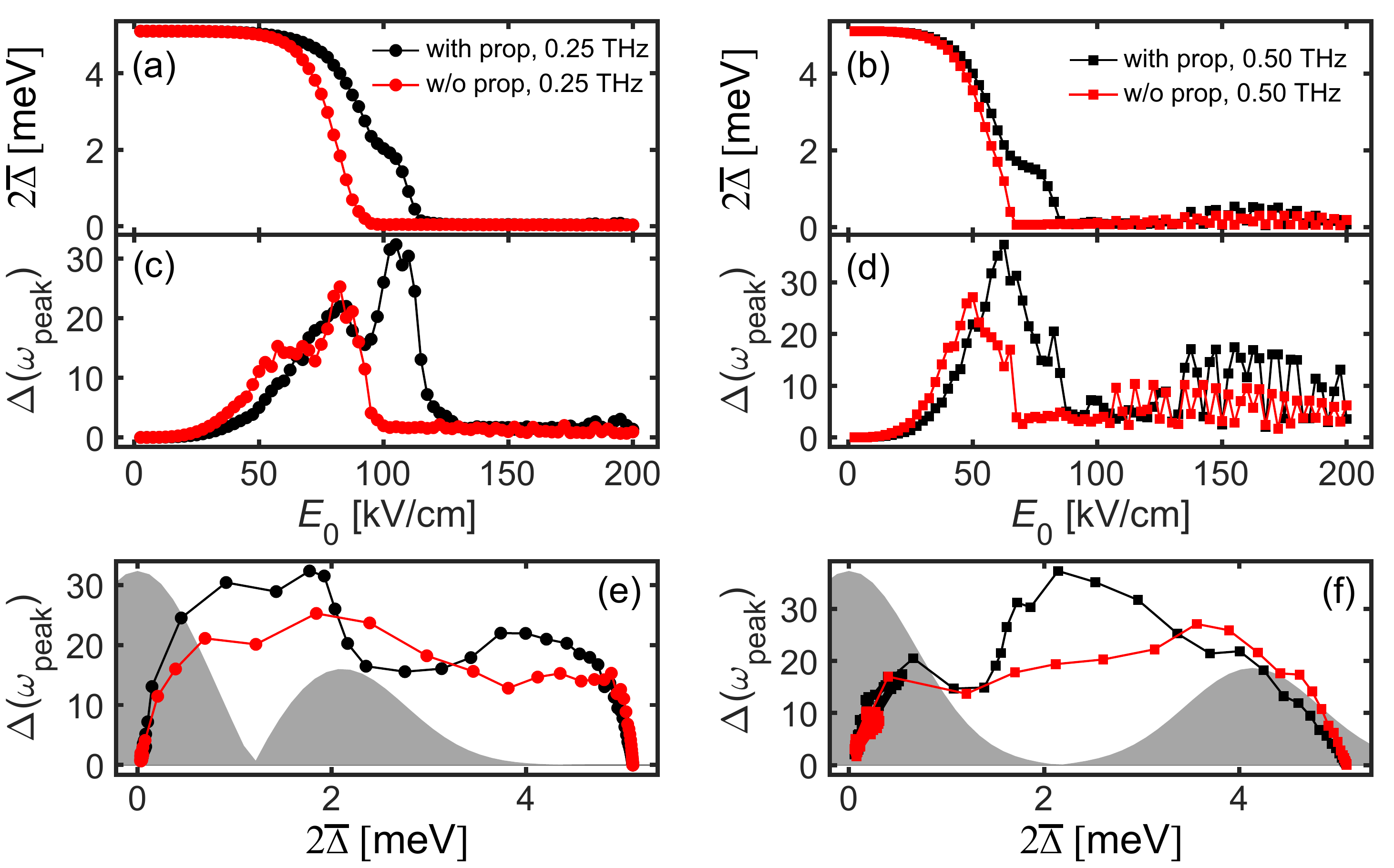}
		\caption{Effect of electromagnetic propagation on collective modes. (a) Electric field strength dependence of the average value of the order parameter in the non-equilibrium state, $\bar{\Delta}$, for a calculation with (black line) and without propagation effects (red line); the SC system has been excited with a 0.25~THz pump field. (c) The corresponding electric field strength dependence of the oscillation amplitude of the main peak in $\Delta(\omega)$ spectra, $\Delta(\omega_\textbf{peak})$. (e) $\Delta(\omega_\textbf{peak})$ as a function of $2\bar{\Delta}$; the Fourier transform of $E^2_\mathrm{THz}(t)$ is shown as a shaded area. (b), (d), (f) The corresponding results for a 0.5~THz pump field.}
		\label{fig11} 
\end{figure*}

Finally, Figure~\ref{fig11} demonstrates that electromagnetic propagation effects do not only lead to photogeneration of a dc current
via THz dynamical symmetry breaking, 
but can also be used to amplify  the oscillation amplitude of the different collective modes. 
The latter  is  controlled 
by the carrier wave cycles of oscillation, which are  
tuned here by varying the frequency and duration 
of the applied THz field. 
We can achieve coherent control by 
synchronizing  the cycles of lightwave oscillations 
with the   SC order parameter and QP dynamics.
Figures~\ref{fig11}(a) and (c) present the pump field dependence of $2\bar{\Delta}$ and $\Delta(\omega_\textbf{peak})$ for the full calculation (black line) and the  calculation without electromagnetic propagation effects (red line) for  a 0.25~THz pump pulse. 
Figure~\ref{fig11}(e) shows $\Delta(\omega_\textbf{peak})$ as a function of $2\bar{\Delta}$, together with the Fourier transform of $E^2_\mathrm{THz}(t)$ (shaded area). The corresponding results for a 0.5~THz pump pulse are plotted in Figs.~\ref{fig11}(b), (d), and (f). Electromagnetic propagation enhances the  collective mode oscillations  when the pumping, 
  $E_\mathrm{THz}^2$, is off-resonant with respect to $2\bar{\Delta}$
(Figure~\ref{fig11}(e) and (f)).   In this case, 
 the finite superfluid momentum $\mathbf{p}_\mathrm{S}$ after the pulse leads to a larger blocking region in the anisotropic ${\bf k}$ distribution of pseudo-spins.
The condensate motion thus   results in much stronger suppression of the anomalous expectation values at certain ${\bf k}$ points in the blocking region.
 As a result, the amplitude modes of the SC order parameter are more strongly excited, which produces larger oscillation amplitudes (collective mode amplification).  The situation changes when  $E_\mathrm{THz}^2$ 
oscillates at a frequency  
  close to $2\bar{\Delta}$. Here, 
 the lightwave  field becomes resonant to $2\Delta$ during the order parameter quench dynamics,  such that resonant Higgs mode excitation dominates
  In particular, the oscillation amplitude is reduced when lightwave propagation is included. We conclude that  the Higgs mode can be amplified by lightwave propagation and by tuning the pump frequency.

\section{Conclusions}

In this paper, we developed  a microscopic gauge-invariant density matrix approach and used it  to study the non-adiabatic nonlinear dynamics of superconductors driven by lightwave electric fields with  few cycles 
of oscillations. In particular, we generalized the Anderson pseudo-spin precession models used in the literature by non-perturbatively including the Cooper pair's center-of-mass motion 
and the condensate spatial variations. We also 
 extended  previous  SC transport theories by including the  non-perturbative coupling of the 
lightwave oscillating strong field  determined by Maxwell's equations
and the nonlinear photocurrent, 
which we showed can break inversion symmetry after the pulse, thus  
leading to gapless non-equilibrium SC and new collective modes. 
 The obtained  gauge-invariant SC Bloch equations, 
together with Maxwell's wave equation, describe the nonlinear dynamical interplay between  lightwave acceleration of the  Cooper-pair condensate, Anderson pseudo-spin nonlinear precession, 
QP Rabi oscillations as well as population inversion,  
spatial dependence, and electromagnetic pulse propagation effects. 
Our theory allows us to treat both amplitude and phase dynamics, driven during cycles of lightwave oscillations, in a gauge-invariant way. 
Such theory  can be extended to treat topological phase ultrafast dynamics.  

We have applied the above comprehensive 
model 
to demonstrate that coherent nonlinearities driven by realistic few-cycle 
THz laser pulses, together with lightwave propagation effects inside a  nonlinear SC thin film system, can photogenerate a nonlinear supercurrent with a dc component. Such  nonlinear supercurrent breaks the equilibrium inversion symmetry of the SC system, which can be detected experimentally via  high-harmonic generation at equilibrium-forbidden frequencies, 
 formation of gapless SC non-equilibrium phases, and 
 Rabi--Higgs collective modes with amplitude amplification. 
The  above  nonlinear effects can be  tuned, e.g.,  by adjusting the thickness of the SC film and the cycles of lightwave oscillation. We have also shown that THz driven Rabi--Higgs flopping and population inversion for sufficiently strong fields 
 can selectively excite and coherently control different classes of collective  modes of the SC order parameter.  More specifically,  we have shown that, with  lightwave condensate acceleration, one can access, in the extreme nonlinear excitation regime, damped harmonic and anharmonic order parameter amplitude oscillations, persisting oscillations, and an overdamped phase. 
  Differences from quantum quench of the SC order parameter studied before arise since the THz electric field  breaks inversion symmetry of electron and hole distributions due to lightwave acceleration of the condensate during cycles of oscillation. 
We thus obtain order-parameter oscillations with multiple frequencies, 
leading to controllable and broad high harmonic generation.  In particular, 
the lightwave-driven damped anharmonic oscillating mode (phase III) and persisting oscillations (phase II) modify the light emission spectrum by producing sideband generation around the fundamental harmonic. 
Finally we have demonstrated that lightwave propagation inside the SC film can significantly amplify the collective mode oscillations.
Such amplification also occurs by driving phase II of persisting order parameter oscillations, i.e., quantum fluctuations due to 
synchronized Rabi oscillations 
at the threshold field for  quenching  the SC order parameter to zero.

The theoretical approach presented here is not restricted to BCS superconductors with a single order parameter. It can be extended, for example, to study the THz-driven non-equilibrium dynamics in multi-band superconductors, in SC systems with multiple coupled order parameters, such as iron-based superconductors~\cite{Patz2014,Patz2017}, 
or in $d$-wave or topological SCs that can be tuned via ${\bf p}_\mathrm{S}(t)$. 
In this connection, one expects to see a rich spectrum of non-equilibrium phases and phase/amplitude collective modes in the non-equilibrium SC dynamics and the extreme nonlinear 
optics regime, to be explored elsewhere. As possible new directions, the derived Bloch--Maxwell equations can be applied to study lightwave propagation effects in  SC  metamaterials, as well as to analyze and predict new multi-dimensional THz coherent nonlinear spectroscopy experiments in superconductors and topological materials. 
For example, in SCs, such experiments provide  a way to distinguish between charge-density fluctuations and collective mode signatures,  study quantum interference and nonlinear wave-mixing effects in quantum states, and generate light-controlled 
collective mode hybridization. Topological order
also leads to analogous effects   determined by the quantum mechanical phase, to be studied elsewhere. 
Phase dynamics in SCs can play an important role 
when spatial variations are considered. 
We conclude  that  THz dynamical symmetry breaking during cycles of coherence oscillations is a powerful concept for addressing quantum sensing and 
coherent control of different quantum materials~\cite{Li2013,Patz2015,2009_PRL_KAPET,chovan-prl, chovan-prb, Lingos2015,Liu2020b} and topological phase transitions~\cite{Luo2019,Yang2020} at the ultimate sub-cycle speed limit necessary for lightwave quantum electronics and magneto-electronics.   

\appendix

\section{Radiative damping}
\label{app:damping}

To study the radiative damping predicted by our theory, we express the density matrix $\tilde{\rho}(\mathbf{k})$ in terms of the Anderson pseudo-spins at each ${\bf k}$ point,
\begin{align}
	\tilde{\rho}(\mathbf{k})=\sum_{n=0}^3 \tilde{\rho}_n(\mathbf{k})\sigma_n\,,
\end{align}
where $\sigma_n$ are the Pauli spin matrices and 
\begin{align}
	&\tilde{\rho}_0=\frac{\tilde{\rho}_{1,1}(\mathbf{k})+\tilde{\rho}_{2,2}(\mathbf{k})}{2}\,,\quad \tilde{\rho}_1=\frac{\tilde{\rho}_{1,2}(\mathbf{k})+\tilde{\rho}_{2,1}(\mathbf{k})}{2}\,, \nonumber \\
	&\tilde{\rho}_2=\mathrm{i}\frac{\tilde{\rho}_{1,2}(\mathbf{k})-\tilde{\rho}_{2,1}(\mathbf{k})}{2}\,,\quad 
	\tilde{\rho}_3=\frac{\tilde{\rho}_{1,1}(\mathbf{k})-\tilde{\rho}_{2,2}(\mathbf{k})}{2}\,,
\end{align}
are the components of the Anderson pseudo-spin. The equations of motion of the pseudo-spin components are 
\begin{align}
\label{eq:eom_ps}
	&\frac{\partial}{\partial t}\tilde{\rho}_0(\mathbf{k})=-e\mathbf{E}\cdot\nabla_\mathbf{k}\tilde{\rho}_3(\mathbf{k})-2|\Delta|\sum_{n=0}^\infty \frac{(\mathbf{p}_\mathrm{S}\cdot\nabla_\mathbf{k})^{2n+1}}{(2n+1)!}\tilde{\rho}_2(\mathbf{k})\,, \nonumber \\
	&\frac{\partial}{\partial t}\tilde{\rho}_1(\mathbf{k})=\left[\xi(\mathbf{k}-\mathbf{p}_\mathrm{S})+\xi(\mathbf{k}+\mathbf{p}_\mathrm{S})-2\,\mu_\mathrm{eff}-2\mu_\mathrm{F}\right]\tilde{\rho}_2(\mathbf{k})\,, \nonumber \\
	&\frac{\partial}{\partial t}\tilde{\rho}_2(\mathbf{k})=-\left[\xi(\mathbf{k}-\mathbf{p}_\mathrm{S})+\xi(\mathbf{k}+\mathbf{p}_\mathrm{S})-2\,\mu_\mathrm{eff}-2\mu_\mathrm{F}\right]\tilde{\rho}_1(\mathbf{k}) \nonumber \\
	&\qquad +2|\Delta|\sum_{n=0}^\infty\left[\frac{(\mathbf{p}_\mathrm{S}\cdot\nabla_\mathbf{k})^{2n}}{(2n)!}\tilde{\rho}_3(\mathbf{k})-\frac{(\mathbf{p}_\mathrm{S}\cdot\nabla_\mathbf{k})^{2n+1}}{(2n+1)!}\tilde{\rho}_0(\mathbf{k})\right]\,, \nonumber \\
	&\frac{\partial}{\partial t}\tilde{\rho}_3(\mathbf{k})=-e\mathbf{E}\cdot\nabla_\mathbf{k}\tilde{\rho}_0(\mathbf{k})+2|\Delta|\sum_{n=0}^\infty\frac{(\mathbf{p}_\mathrm{S}\cdot\nabla_\mathbf{k})^{2n}}{(2n)!}\tilde{\rho}_2(\mathbf{k})\,.
\end{align}
We then linearize the equations of motion (\ref{eq:eom_ps}) with respect to deviations from equilibrium yielding
\begin{align}
\label{eq:eom-lin}
	&\frac{\partial}{\partial t}\delta\tilde{\rho}_0(\mathbf{k})=-e\mathbf{E}\cdot\nabla_\mathbf{k}\tilde{\rho}_3^{(0)}(\mathbf{k})\,,\nonumber \\
	&\frac{\partial}{\partial t}\delta\tilde{\rho}_1(\mathbf{k})=2\xi(\mathbf{k})\delta\tilde{\rho}_2(\mathbf{k})\,, \nonumber \\
	&\frac{\partial}{\partial t}\delta\tilde{\rho}_2(\mathbf{k})=2\xi(\mathbf{k})\delta\tilde{\rho}_1(\mathbf{k})-2\delta\mu_\mathrm{eff}\tilde{\rho}_1^{(0)}\nonumber \\ 
	&\qquad+2\Delta_0\delta\tilde{\rho}_3(\mathbf{k})+2\delta\Delta\tilde{\rho}_3^{(0)}(\mathbf{k})\,,\nonumber \\
	&\frac{\partial}{\partial t}\delta\tilde{\rho}_3(\mathbf{k})=2\Delta_0\delta\tilde{\rho}_2(\mathbf{k})\,,
\end{align}
where 
\begin{align}
\label{eq:rho_ps}
&\delta\tilde{\rho}_n(\mathbf{k})=\tilde{\rho}_n(\mathbf{k})-\tilde{\rho}^{(0)}_n(\mathbf{k})\,, \quad 
\delta\mu_\mathrm{eff}=\frac{\mathrm{i}}{2}\frac{\partial}{\partial t}\theta+e\phi+\delta\mu_\mathrm{F}\,,\nonumber \\ &\delta\delta\mu_\mathrm{F}=g\sum_\mathbf{k}\delta\tilde{\rho}_3(\mathbf{k})\,, \quad \delta\Delta=\Delta(t)-\Delta_0\,,
\end{align}
with equilibrium pseudo-spin components
\begin{align}
	&\tilde{\rho}^{(0)}_0(\mathbf{k})=1\,, \quad \tilde{\rho}^{(0)}_1(\mathbf{k})=\frac{\Delta_0}{\sqrt{\xi(\mathbf{k})^2+\Delta_0^2}}\,, \nonumber \\
	&\tilde{\rho}^{(0)}_2(\mathbf{k})=0\,, \quad \tilde{\rho}^{(0)}_3(\mathbf{k})=-\frac{\xi(\mathbf{k})}{\sqrt{\xi(\mathbf{k})^2+\Delta_0^2}}\,.
\end{align}
Here we applied perturbation theory with respect to linear order in the electric field $\mathbf{E}$, i.~e. we have neglected all contributions of order $\mathcal{O}(\mathbf{E}^2)$ and higher. We next Fourier transform Eq.~(\ref{eq:eom-lin}) and insert the result into the Fourier transformation of the current (\ref{eq:current}). By combining the result with the Fourier transformation of Eq.~(\ref{eq:wave_eq-sol}), we find
\begin{align}
	&|J(\omega)|^2=4n^2\varepsilon_0^2\Gamma^2\frac{|\mathbf{E}_0(\omega)|^2}{\omega^2+\Gamma^2}\,,
\end{align}
where we introduced the radiative coupling constant
\begin{align}
\label{eq:gamma}
	&\Gamma=\frac{e^2}{S\hbar^2}\frac{1}{2n\varepsilon_0 c}\sum_\mathbf{k}\frac{\partial}{\partial k_x}\xi(\mathbf{k})\frac{\partial}{\partial k_x}\tilde{\rho}_3^{(0)}(\mathbf{k})\,,
\end{align}
after assuming that the applied electric field $\mathbf{E}_0$ is polarized in $x$-direction. The self-consistent coupling between the photoexcited current and the lightwave field thus induces a radiative damping, which is given by Eq.~(\ref{eq:gamma}) in linear order perturbation theory.

The transformation from particle space to quasi-particle space is performed using the unitary Bogoliubov transformation
\begin{align}
\label{eq:qp_trafo}
	&\rho^\mathrm{qp}(\mathbf{k})=\mathcal{U}_\mathbf{k}\,\tilde{\rho}(\mathbf{k})\,\mathcal{U}_\mathbf{k}^\dagger\,, \quad \mathcal{U}_\mathbf{k}=
\begin{pmatrix}
u_\mathbf{k} & v_\mathbf{k} \\
-v_\mathbf{k} & u_\mathbf{k}
\end{pmatrix}\,,
\end{align}
with coherence factors
\begin{align}
\label{eq:coh}
	&u_\mathbf{k}=\sqrt{\frac{1}{2}\left(1+\frac{\varepsilon_{\mathbf{k}}}{E_\mathbf{k}}\right)}\,, \quad v_\mathbf{k}=\sqrt{\frac{1}{2}\left(1-\frac{\varepsilon_{\mathbf{k}}}{E_\mathbf{k}}\right)}\,,
\end{align}
where 
\begin{align}
\label{eq:coh2}
	&\varepsilon_\mathbf{k}=\frac{\xi(\mathbf{k}+\mathbf{p}_\mathrm{S}/2)+\xi(\mathbf{k}-\mathbf{p}_\mathrm{S}/2)}{2}\,,\quad E_\mathbf{k}=\sqrt{\varepsilon_\mathbf{k}^2+|\Delta|^2}\,.
\end{align}
Here we have chosen an instantaneous quasi-particle basis with time-dependent coherence factors (\ref{eq:coh}), which diagonalizes the time-dependent homogeneous Hamiltonian. The corresponding quasi-particle energies are given by
\begin{align}
\label{eq:Eqp}
	&E^\mathrm{qp}_{\mathbf{k},\pm}=\frac{\xi(\mathbf{k}+\mathbf{p}_\mathrm{S}/2)-\xi(\mathbf{k}-\mathbf{p}_\mathrm{S}/2)}{2}\pm E_\mathbf{k}\,.
\end{align}

\begin{acknowledgments} 
Theory work at the University of Alabama, Birmingham was supported by the US Department of Energy Office of Science, Basic Energy Sciences 
 under contract\#DE-SC0019137 (M.M and I.E.P). It was also made possible in part by a grant for high performance computing resources and technical support from
the Alabama Supercomputer Authority. J.W. was supported by the Ames Laboratory,  US Department of Energy, Office of Science,
Basic Energy Sciences, Materials Science and Engineering Division under contract No. DEAC02-07CH11358.

\end{acknowledgments}


%

\end{document}